\documentclass[
twocolumn,  
secnumarabic,
amssymb, 
nobibnotes,
aps, 
prd,
superscriptaddress,
nofootinbib,
]{revtex4-1}
\usepackage{algorithm}

\newcommand{\diff}{{\rm d}}

\usepackage[ddmmyyyy]{datetime} 
\usepackage{amsmath} 
\usepackage[dvipsnames]{xcolor}
\usepackage{cancel} 
\usepackage{graphicx}
\usepackage{comment} 
\usepackage{soul,ulem} 
\usepackage{eqnarray}
\soulregister\cite7
\soulregister\ref7 
\soulregister\eqref7 
\usepackage{bbding}
\usepackage{pifont}
\usepackage{wasysym}
\usepackage{amssymb}
\usepackage{textcomp} 
\usepackage{marginnote} 
\usepackage[colorinlistoftodos
, textsize=tiny]{todonotes} 
\usepackage{textcomp} 
\usepackage{lineno} 
\usepackage{hyperref}
\usepackage{cleveref}
\hypersetup{
    colorlinks=true,
    citecolor= teal, 
    linkcolor= [rgb]{0.1,0.7,0.1},
    filecolor=magenta,      
    urlcolor=[rgb]{0.1,0.5,0.5},
    urlbordercolor=red
}

\newcommand*\aap{Astron.~Astrophys.}

\newcommand*\jcap{J. Cosmology Astropart. Phys.}

\newcommand*\mnras{Mon.~Not.~Roy.~Astron.~Soc.}

\newcommand{\M}{\ensuremath{M_{500}}}
\newcommand{\R}{\ensuremath{R_{500}}}
\newcommand{\C}{\ensuremath{c_{500}}}

\newcommand{\Psz}{\ensuremath{P_{\rm SZ}}}
\newcommand{\Px}{\ensuremath{P_{\rm X}}}

\newcommand{\phitwo}{\ensuremath{\phi_{\infty, 2}}}
\newcommand{\btwo}{\ensuremath{\beta_{2}}}

\newcommand{\fofr}{\text{$f(R)$} }
\newcommand{\fR}{|f_{R0}| }
\defcitealias{Ettori:2018tus}{E19}

\begin{document}

\title{ Constraining Chameleon screening using galaxy cluster dynamics}
\author{Yacer Boumechta}
\email{yboumech@sissa.it}
\affiliation{SISSA-International School for Advanced Studies, Via Bonomea 265, 34136 Trieste, Italy}
\affiliation{ICTP-The Abdus Salam International Centre for Theoretical Physics, Strada Costiera 11, 34151 Trieste, Italy  }
\affiliation{IFPU, Institute for Fundamental Physics of the Universe, via Beirut 2, 34151 Trieste, Italy}
\affiliation{INFN, Sezione di Trieste, Via Valerio 2, I-34127 Trieste, Italy}

\author{Balakrishna S. Haridasu}
 \email{sandeep.haridasu@sissa.it}
 \affiliation{SISSA-International School for Advanced Studies, Via Bonomea 265, 34136 Trieste, Italy}
 \affiliation{INFN, Sezione di Trieste, Via Valerio 2, I-34127 Trieste, Italy}
 \affiliation{IFPU, Institute for Fundamental Physics of the Universe, via Beirut 2, 34151 Trieste, Italy}
 \author{Lorenzo Pizzuti}
 \email{pizzuti@fzu.cz}
 \affiliation{CEICO, Institute of Physics of the Czech Academy of Sciences, Na Slovance 2, 182 21 Praha 8, Czechia}
\author{Minahil Adil Butt}
\email{mbutt@sissa.it}

\author{Carlo Baccigalupi}
\email{bacci@sissa.it}
\affiliation{SISSA-International School for Advanced Studies, Via Bonomea 265, 34136 Trieste, Italy}
 \affiliation{IFPU, Institute for Fundamental Physics of the Universe, via Beirut 2, 34151 Trieste, Italy}
 \affiliation{INFN, Sezione di Trieste, Via Valerio 2, I-34127 Trieste, Italy}
\author{Andrea Lapi}
\email{lapi@sissa.it}
\affiliation{SISSA-International School for Advanced Studies, Via Bonomea 265, 34136 Trieste, Italy}
 \affiliation{IFPU, Institute for Fundamental Physics of the Universe, via Beirut 2, 34151 Trieste, Italy}
 \affiliation{INFN, Sezione di Trieste, Via Valerio 2, I-34127 Trieste, Italy}
 \affiliation{IRA-INAF, Via Gobetti 101, 40129 Bologna, Italy}

\begin{abstract}

We constrain the Chameleon \textit{screening} mechanism in galaxy clusters, essentially obtaining limits on the coupling strength $\beta$ and the asymptotic value of the field $\phi_{\infty}$. For this purpose, we utilized a collection of the 9 relaxed galaxy clusters within the X-COP compilation in the redshift range of $z \le 0.1$. We implement the formalism assuming an NFW mass profile for the dark matter density and study the degeneracy present between the mass $\M$ and the chameleon coupling with a high degree of improvement in the constraints for excluded parameter space. We recast our constrain to an upper limit on the scalaron field in \fofr sub-class of models of $|f_{R0}|\le 9.2\times 10^{-6}$, using all the nine clusters and $|f_{R0}|\le 1.2\times 10^{-5}$ using only 5 clusters with WL priors taken into account, at a $95\%$ confidence level. These bounds are consistent with existing limits in the literature and tighter than the constraints obtained with the same method by previous studies.

\end{abstract}

\maketitle


\section{Introduction}
\label{sec:Introduction}

{Most cosmological observations \cite{Adam:2015rua, Ade:2015xua} can be explained, to a high degree of precision, within the framework of General Relativity (GR). In particular, adding a phenomenological cosmological constant ($\Lambda$) \cite{Weinberg:1988cp, Bull:2015stt} to the Einstein field equations can account for the late-time acceleration of the universe \cite{Haridasu:2017lma}. Despite its success in reproducing a wide range of datasets (e.g. \cite{Planck2020}), the Concordance $\Lambda$CDM model is not still able to provide a physically-acceptable motivation for the nature of the cosmological constant. For such reason, in the last decades, alternative viewpoints have been proposed} by allowing additional degrees of freedom that could elucidate the dark energy as a dynamic field (for instance quintessence \cite{Caldwell:1997ii}), or Modify GR \cite{Tsujikawa:2010zza,Nojiri:2010wj, Clifton:2011jh} so that it can account for the Dark energy effects \cite{Tsujikawa:2010zza}. 

{One of the most popular wide class of Modified Gravity (MG) models is represented by the framework of scalar-tensor theories \cite{Langlois:2015cwa, Crisostomi:2016czh,BenAchour:2016fzp, Motohashi:2016ftl}, where a scalar degree of freedom is added to the gravitational field. The presence of such scalar field provides an additional contribution to the gravitational force \cite{Khoury:2013yya, Burrage:2017shh}, leaving (in principle) detectable imprints on the formation and evolution of cosmic structures\cite{Brax:2005ew, Brax:2004qh, Brax:2013mua,Banerjee:2008rs}. 
This new interaction should be suppressed at small scales and high density regions in order to match the tight constraints of GR. Depending on the implementation of this screening mechanism, the effect of the new (fifth) force on matter density perturbations can be significantly different.}

{A particularly interesting subset of scalar-tensor models is the Chameleon field theory\cite{Khoury:2003rn,Faulkner:2006ub,Navarro:2006mw}, where the additional scalar field couples non-minimally to the matter and introduces a fifth force \cite{Khoury:2003aq}. The screening is achieved by working on the potential associated to the field, making the effective mass very large in high density regions such that the force is suppressed. The modification of the gravitational interaction becomes important at a large distance from the center of a matter distribution \cite{Khoury:2003aq}.
When the fifth force is active, it affects the motion of non-relativistic objects such as galaxies and hot diffuse gas in galaxy clusters. In particular, the presence of the Chameleon field changes the relation between pressure and gravitational potential of the hot Intra-Cluster-Medium (ICM) of a cluster \cite{Terukina:2013eqa,Terukina12,Lombriser:2013eza,Tamosiunas:2021kth}}. 

Two main parameters construct the chameleon field model in a galaxy cluster: first one is $\beta$, which is the coupling constant between the Chameleon field and matter density ans the latter, $\phi_\infty$  which is the intensity of the field at a larger distance away from the cluster. Under reasonable assumptions (e.g. \cite{Terukina:2013eqa}) these two parameters describe the modification of gravity completely. Also, the case of $\beta=\sqrt{1/6}$ within the Chameleon field scenario describes an $f(R)$ theory \cite{Starobinsky:2007hu,Oyaizu:2008tb}. 

We consider that the total mass distribution of a galaxy cluster can be parametrized by a Navarro–Frenk–White (NFW) density model \cite{Navarro:1995iw, Wyithe:2000si, Zavala:2006de, Matos:2003nb, Dehghani:2020cvl, Asano:1999ti}; 
under the assumption of hydrostatic equilibrium, the total gravitational potential of the cluster will affect the pressure of the hot gas \cite{Terukina:2013eqa}.  In this paper, we implement the formalism presented in \cite{Terukina:2013eqa}to the XMM-Newton
Cluster Outskirts Project X-COP data products \cite{Ettori:2016kll, Ettori:2018tus, Eckert:2016bfe, Ghirardini:2018byi}, which consists of 12 clusters with well-observed X-ray emission and high signal to noise ratio in the Planck
Sunyaev-Zel’dovich (SZ) survey \cite{Planck:2015lwi}, essentially providing both ICM temperature and pressure data over the large radial range of $0.2\, \rm{Mpc} \le r \le 2 \,  \rm{Mpc}$.


We use the X-ray temperature, SZ pressure, and electron density data to derive constraints on the Chameleon parameters $\beta$ and $\phi_\infty$ by performing a Monte-Carlo-Markov-Chain (MCMC) analysis. In our computation, we use a self-contained prior on the mass profile which is referred to as "internal mass prior" (see  \ref{sec:Results}), that can be justified by analyzing the effect of adding a weak lensing mass estimation available for five clusters \cite{Herbonnet:2019byy}. Indeed, due to the conformal structure of the chameleon model, gravitational lensing analyses are not affected by the fifth force; thus information derived from lensing can be used to break the degeneracy among model parameters. We also discuss the effect of fixing the electron density parameters in the analysis, which are not correlated to the MG parameters.

The paper is organized as follows: in \Cref{Sec:Modell} we construct our model for the Chameleon field and show the solution of this field as applied to  galaxy cluster with the assumption of the NFW profile, and at the end of the section we discuss the effect of the modification induced by the presence of Chameleon field on the Hydrostatic pressure.  In \Cref{sec:data} we present briefly  the X-COP data and then construct the likelihood that we will use with the MCMC analysis to generate the chains that constrain our parameter space. In \Cref{sec:Results} we present our results and discuss them in detail while comparing our constraints with the ones obtained by other galaxy clusters' analyses (e.g. \cite{Terukina:2013eqa,Wilcox:2015kna}). Finally, we further derive our main conclusions in \Cref{sec:conclusions}.

\section{Modeling}
\label{Sec:Modell}
In this section, we briefly review the framework of Chameleon screening mechanism, highlighting the main features relevant for our analysis. 
\subsection{Screening Mechanism }
The Lagrangian of the theory includes the usual Einstein- Hilbert Lagrangian plus the scalar field, in addition to the Standard Model fields coupled minimally to gravity
\cite{Khoury:2003aq,Zaregonbadi:2022lpw,Ivanov:2016ucz,Kraiselburd:2015vyf,Tsujikawa:2009yf},

\begin{equation}
\label{eqn:1}
L=\frac{M^2_\text{Pl}}{2}R+L_{m}(\tilde{g}_{\mu\nu},\psi)+L_{\phi}, 
\end{equation}
where  $L_{\phi}=-\frac{1}{2}(\partial\phi)^{2}-V(\phi)$ and $ M_\text{Pl}=\frac{1}{\sqrt{8\pi G}}$; the Standard Model fields are represented by $\psi$, and  $\tilde{g}_{\mu\nu}=A^{2}(\phi)g_{\mu\nu}$. In the quasi-static approximation, the equation of motion for the field $\phi$ can be written as \cite{Kase:2013uja},

\begin{equation}
\label{eqn:2}
    \nabla^{2}\phi=V'(\phi)-\frac{A'(\phi)}{A(\phi)}T.
\end{equation}
Here the $'$ represents the derivative with respect to $\phi$ and $T$ is the trace of the stress-energy tensor of the standard model field $\psi$. 
One can notice that the Chameleon field dynamics is sourced by the trace of the stress-energy tensor as is shown in  \Cref{eqn:2}; the field values depend on the matter component and thus the field behaves in different ways for different matter distributions. We denote $\frac{A'(\phi)}{A(\phi)}=\frac{\beta}{M_\text{Pl}}$ which is going to be a constant in the current formalism,  and here $\beta$ is the coupling factor between the field $\phi$ and the stress-energy tensor $T$. Finally we consider only pressureless non-relativistic matter fields, which implies $T=-\rho_{m}$.

Therefore we can write,  
\begin{equation}
\label{eqn:3}
    \nabla^{2}\phi=V_{\rm eff}'(\phi),
\end{equation}
where 
\begin{equation}
\label{eqn:4}
    V_{\rm eff}(\phi)=V(\phi)+\frac{\beta\phi}{M_\text{Pl}}\rho_{m}\,.
\end{equation}
{The form of the potential $V(\phi)$ should guarantee that the gravitational effect induced by this field} will be suppressed when we have large matter densities i.e. the field $\phi$ is screened and GR is recovered. {On the other hand, at lower densities, we want the effect of the field to become important}, which will require us to impose that the potential $V(\phi)$ is a decreasing function of $\phi$ \cite{Khoury:2003aq}, {typically a power-law potential $V(\phi)=\Lambda^{4+n}\phi^{-n}$, where $\Lambda$ and $n$ are constants.} 

In the region where $\phi$ is un-screened, an additional fifth force is induced by the gradient of the Chameleon field,
\begin{equation}
\label{eqn:5}
F_\phi=-\frac{\beta}{M_\text{Pl}}\nabla \phi\,,
\end{equation}
{providing an additional contribution to the Newtonian potential while retaining hydrostatic equilibrium assumption in chameleon gravity.}

\subsection{Chameleon field in Cluster of Galaxies}
In the following analysis, we assume that the total matter density distribution within the galaxy cluster can be modeled as a NFW profile \cite{Navarro:1995iw},

\begin{equation}
\label{eqn:6}
\rho(r) = \frac{\rho_{s}}{r/r_{s}(1+r/r_{s})^{2}}\,,    
\end{equation}
where $\rho_\text{s}$ and $r_\text{s}$ are characteristic density and scale radius, respectively. {The NFW model has been shown to provide a good description for simulated DM halos ( see e.g. \cite{Schaller15}) and for real clusters' data in $\Lambda$CDM (e.g. \cite{Hogan2017,Sartoris2020}), while some other works have further suggested that the NFW profile performs well also in  modified gravity scenarios, including chameleon gravity \cite{Lomb12,Wilcox:2016guw,Naik19}.}

We are interested in finding the solution for the chameleon Equation (\ref{eqn:3}) in the presence of a matter density distribution given by Equation (\ref{eqn:6}); in order to do that, we employ the semi-analytical approach followed by e.g. \cite{Terukina:2013eqa}. The idea is that below some radius $r_{c}$, the value of the scalar field at the interior minimizes the effective potential $V_{eff}(\phi)$ which represents the regime where the Chameleon force does not contribute and the solution is obtained by setting $\nabla\phi=0$ in the left-hand side of Equation \eqref{eqn:3}. On the other hand, at larger distances, the potential $V(\phi)$ is negligible and the second term in \eqref{eqn:4} dominates the effective potential. The solution in this regime is obtained by solving  $\nabla^{2}\phi={\beta\phi}\rho_{m}/{M_\text{Pl}}$. Therefore, we obtain the complete semi-analytical solution as, 
\begin{equation}
\label{eqn:7}
\phi(r)=
\begin{cases}
\phi_{s}\left[r/r_{s}(1+r/r_{s})^{2}\right]=\phi_{\rm int}\simeq0 & r<r_{c}\\
-\frac{\beta\rho_{s}r_{s}^{2}}{M_\text{Pl}}\frac{\ln(1+r/r_{s})}{r/r_{s}}-\frac{C}{r/r_{s}}+\phi_{\infty}=\phi_{\rm out} & r>r_{c}\,.
\end{cases}
\end{equation}

In the above equation, $\phi_{s}$ is a constant which depends on the characteristic density and the parameters of the potential $V(\phi)$. The integration constant $C$ and  the radius $r_c$ can be specified by imposing the continuity of the solution and its first derivative at $r=r_c$. Thus we have \cite{Terukina:2013eqa},
\begin{equation}
\label{eqn:8}
1+\frac{r_{c}}{r_{s}} \simeq \frac{\beta\rho_{s}r_{s}^{2}}{M_\text{Pl}\phi_{\infty}}  
\end{equation}
\begin{equation}
\label{eqn:9}
    C \simeq -\frac{\beta \rho_s r_s^2}{M_{\rm Pl}}\ln(1+r/r_s) +\phi_\infty r/r_c
\end{equation}
{The \textit{screening} radius $r_c$ represents the transition below which the Chameleon field is screened, and as shown in \Cref{eqn:8}, it is completely determined by the other parameters of the model. In particular, the screening radius is strongly dependent on the mass of the cluster $M_{500}\propto r_s^3\rho_s$ (see \Cref{eqn:18,eqn:19}). Which implies that in massive clusters the screening mechanism tends to be very efficient, while the fifth force is more active in lower mass halos (e.g. \cite{Pizzuti:2020tdl}).}

\subsection{$\fofr$ analogy with Chameleon field}

$f(R)$ gravity \cite{buchadal70} is one of the most investigated alternatives of GR at cosmological level; in this class of models, the Einstein-Hilbert action is modified by adding a generic function of the Ricci scalar:
\begin{equation}
\label{eqn:10}
    S=\int\text{d}^4x\sqrt{-g}\frac{1}{16\pi G}\left[R+f(R)\right]+S_m[\psi_i,g_{\mu\nu}]\,.
\end{equation}
The functional form of $f(R)$ can be chosen in such a way that the background $\Lambda$CDM expansion history is reproduced as close as desired (see e.g. \cite{Hu:2007nk}). The derivative of the function $f_R ={\partial f_{R}}/{\partial R} $ plays a role of a dynamical scalar field which, under certain conditions can be conformally recasted into a scalar-tensor model exhibiting chameleon screening (see e.g. \cite{Brax:2008}).
This is possible in particular if the scalar field $f_R ={\partial f_{R}}/{\partial R} $, called \textit{scalaron}, has a positive large effective mass at high curvature\cite{Song:2006ej}. 

The field equation for $f_R$ is  \cite{Hu:2007nk}
\begin{equation}
    \label{eqn:11}
    \Box f_{R}=\frac{\partial V_{\rm eff}(f_{R})}{\partial f_{R}}\,,
\end{equation}
which is analogous to Equation  \ref{eqn:3}  with the replacement:
\begin{equation}
\label{eqm:12}
    f_R=\exp\left(-\frac{2\beta\phi}{M_\text{Pl}}\right)\,,
\end{equation}
and $\beta =\sqrt{1/6}$ \cite{Terukina:2013eqa, Pizzuti:2022ynt}. The value of the scalar field for the background today $f_{R0}=\bar{f}_R(z=0)$ is proportional to the present value of the chameleon field at infinity as $ f_{R0}=-\sqrt{\frac{2}{3}}\frac{\phi_\infty}{M_\text{Pl}}$.

In the last decades, several works have placed constraints on $f(R)$ gravity using different probes, both at astrophysical (e.g. \cite{Jain:2012tn, Jain:2015edg,Pretel2020}) and at cosmological (e.g \cite{Terukina:2013eqa, Wilcox:2015kna,Raveri:2014cka,Cataneo:2016iav, Pizzuti17,Perico19} ) scales. Currently, the most stringent bounds on the scalaron are of the order of $f_{R0}\lesssim 10^{-7}$, for particular choices of $f(R)$, from galaxy rotation curves \cite{Naik19}, while cosmological analyses limit the background field to be $f_{R0}\lesssim 10^{-6}$ (e.g. \cite{Xu:2015}). 

\subsection{Hydrostatic Equilibrium}
For a spherical system that contains gas with pressure $P$ and density $\rho_{g}$, the hydrostatic equilibrium equation is given by,
\begin{equation}
\label{eqn:13}
\frac{1}{\rho_{g}}\frac{\diff P(r)}{\diff r}=-\frac{GM(<r)}{r^{2}}\,,
\end{equation}
where $M(r)$ is the mass enclosed within the radius $r$, and the above equation represents the balance between the force induced by the gas pressure and the gravitational force. However, in the current MG scenario, we have an additional force given by \Cref{eqn:5} induced by the existence of the Chameleon field, which contributes as a new term in the hydrostatic equilibrium equation as \cite{Terukina:2013eqa}, 

\begin{equation}
\label{eqn:14}
    \frac{1}{\rho_{g}}\frac{\diff P(r)}{\diff r}=-\frac{GM(r)}{r^{2}}-\frac{\beta}{M_\text{Pl}}\frac{\diff \phi(r)}{\diff r}\,,
\end{equation}
which upon integration provides 

\begin{equation}
\label{eqn:15}
    P(r)=P_{0}-\mu m_{p}\int_{0}^{r}n_{e}(r)\left[\frac{GM(r)}{r^{2}}+\frac{\beta}{M_\text{Pl}}\frac{\diff\phi(r)}{\diff r}\right] \diff r \,, 
\end{equation}

Where $\mu$ is the mean molecular weight, $P_0$ is an integration constant, i.e, pressure at $r=0$, and $n_e(r)$ is the electron density at radius $r$.
We further assume the electron density to follow the Vikhlinin profile \cite{Vikhlinin:2005mp,SPT:2017ees},
\begin{equation}
\label{eqn:16}
   n_{e}(r)=n_{0}\frac{\left(\frac{r}{r_{1}}\right)^{-\alpha_v/2}\left[1+(\frac{r}{r_{2}})^{\gamma_v}\right]^{-{\epsilon_v}/{2\gamma_v}}}{\left[1+(\frac{r}{r_{1}})^{2}\right]^{3\beta_v/2-\alpha_v/4}}
\end{equation}
where we fix $\gamma_v=3$ as suggested in \cite{Vikhlinin:2005mp}. The electron density profile above thus contains 6 parameters. While the original Vikhlinin profile contains 9 parameters, we have earlier validated that the 6 parameter reduced form is sufficient for the dataset utilized here \cite{Haridasu:2021hzq}. 
\begin{figure}
    \centering
    \includegraphics[scale=0.48]{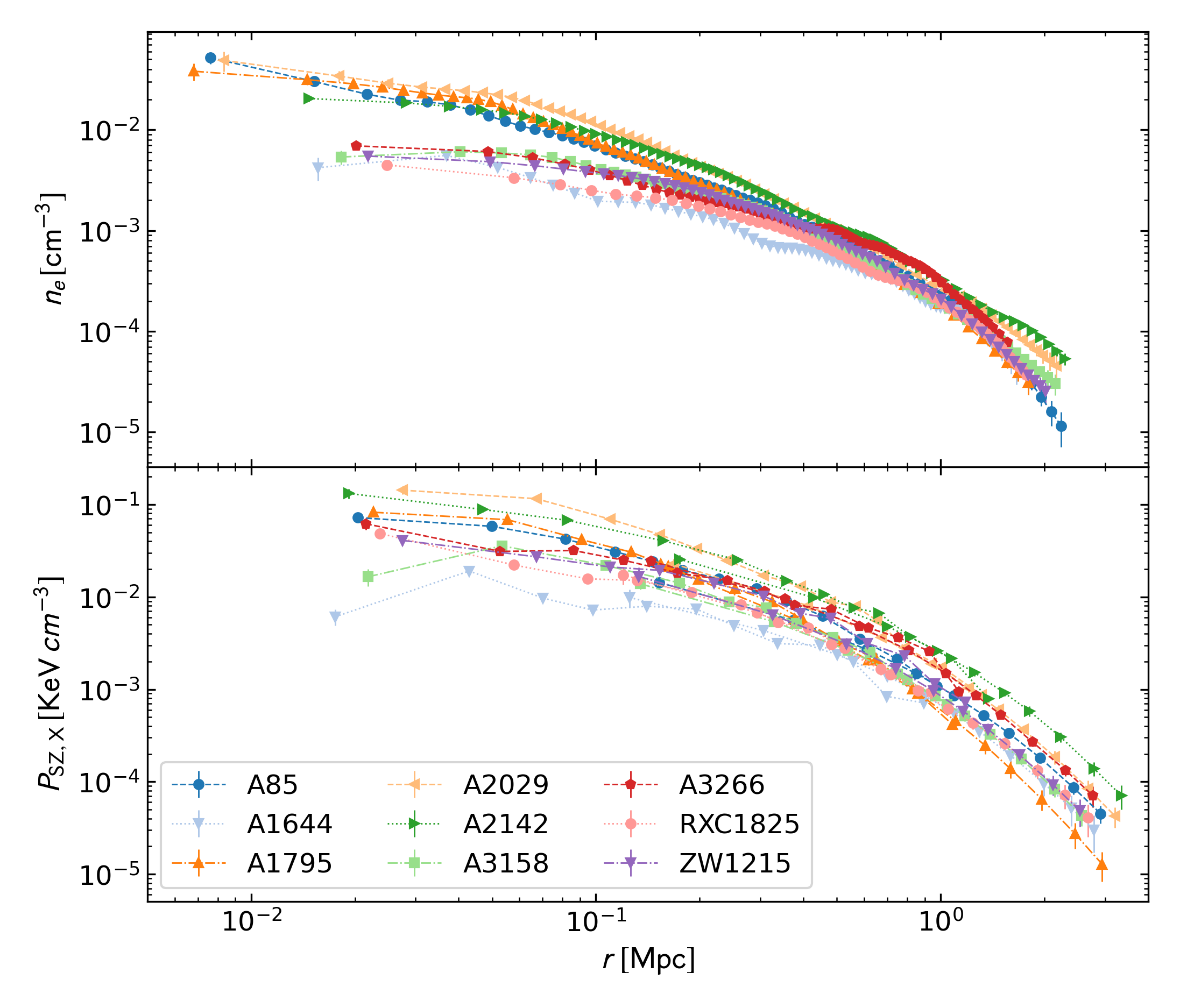}
    \caption{\textit{Top}: Radial profiles of the electron density \cite{Ghirardini:2018byi}. \textit{Bottom}: Pressure data obtained using the Compton effect ($\Psz$) and the X-ray observations ($\Px$) \cite{Ettori:2018tus}. We show the data for all the nine clusters we have utilized in the current analysis. }
    \label{fig:XCOP}
\end{figure}

\section{Data and likelihood}
\label{sec:data}

\subsection{X-COP clusters}
We utilize 9 X-COP clusters \cite{Ettori:2018tus}, following the formalism  utilized in an earlier work \cite{Haridasu:2021hzq, Haridasu:2021yrw}. We keep the current section brief as the utility of the datasets is equivalent to the aforementioned application. {While a total of 12 clusters are available in the X-COP compilation, in the current work we utilize only 9 of them excluding A644, A2255 and A2319. The 3 excluded clusters do no favor the NFW mass profile which is an integral assumption in obtaining the semi-analytical expressions for the field in the formalism adopted here. We however include A1644 which is reported to perform equivalently for NFW and the best-fit Hernquist mass profile.  We defer the study of the effects of mass profile assumptions on the constrains on the screening mechanisms to a later communication.} We show the final datasets of the electron density (\textit{top}), and pressure obtained using both X-ray and SZ methods \cite{Ade:2013skr} (\textit{bottom}) in \Cref{fig:XCOP}.

\subsection{Likelihood}
The complete formalism introduced in \Cref{Sec:Modell} is described by 10 parameters; 2 defining the Chameleon field ($\beta$ and $\phi_{\infty}$), 2 for the NFW profile ($\rho_{s}$ and $r_{s}$), the remaining 6 parameters are from the expression of the electron density given by \Cref{eqn:16}. The individual likelihood ($-2 \ln{\mathcal{L}}$) for the pressure and electron density data are then written as,

\begin{multline}
\label{enq:20}
\chi^2_{P}=
(\textbf{P}_{SZ}^{obs}-\textbf{P}_{SZ})\Sigma_{P}^{-1}(\textbf{P}_{SZ}^{obs}-\textbf{P}_{SZ})^{T}+\ln\left|\Sigma_{p}\right|+\\ \sum_{i}\left[\frac{(P_{X}(r_{i})-P_{X,i}^{obs})^{2}}{\sigma_{P_{X,i}}^{2}+\sigma_{int}^{2}}+\ln(\sigma_{P_{X,i}}^{2}+\sigma_{\rm int}^{2})\right]\,,
\end{multline}

\begin{equation}
    \label{enq:21}
\chi^2_{n_e}=\sum_{i}\frac{(n_{e}(r_{i})-n_{e,i}^{obs})^{2}}{\sigma_{n_{e,i}}^{2}}\,,
\end{equation}
respectively. The total $\chi^2$ function is then the summation individual contributions, upon which we perform the Bayesian analysis and is given by,
\begin{equation}
\label{eqn:17}
-2 \ln{\mathcal{L}^{\rm tot}} \equiv \chi^2 (\Theta_{n_e},\Theta_{\rm NFW},\Theta_{\rm MG})=   \chi^2_{\rm P} + \chi^2_{n_e}
\end{equation}
wherein
$\Theta_{n_e}=\{n_0,r_1,r_2,\alpha_v,\beta_v,\epsilon_v\}$, $\Theta_{\rm NFW}=\{M_{500},c_{500}\}$
and $\Theta_{\rm MG}=\{\phi_{\infty,2},\beta_2\}$.
Refer to \cite{Haridasu:2021hzq, Haridasu:2021yrw}, for further details on the likelihood and the  inclusion of the intrinsic scatter ($\sigma_{\rm int}$) parameter. 

Therefore, we perform a MCMC analysis over a 10-dimensional parameter space  $(n_0, r_1, r_2, \alpha, \beta, M_{500}, c_{500}, \phi_{\infty,2}, \beta_2)$ , where the two parameters $\beta_{2} $ and $ \phi_{\infty,2}$ are compactified functions of $\beta$ and $\phi_{\infty}$, respectively, and are given by $\beta_{2}={\beta}/(1+\beta)$ and $\phi_{\infty,2}=1-\exp(-\phi_{\infty}/10^{-4}M_\text{Pl})$. These new scaled parameters run in the interval $[0,1]$, making the interpretation of the results straightforward. It is also convenient to use $M_{500}$ and $\R$ instead of $\rho_s$ and $r_s$ which are related through the following relations \cite{Terukina:2013eqa}:
\begin{equation}
\label{eqn:18}
    r_s=\frac{1}{c_{500}}\left[\frac{M_{500}}{(4\pi/3)\Delta_c \rho_c}\right]^{1/3}
\end{equation}
\begin{equation}
\label{eqn:19}
    \rho_s=\frac{\M}{4\pi r^3_s} \left(\ln(1+c_{500})-\frac{c_{500}}{c_{500}+1}\right)^{-1}
\end{equation}
where $\C={\R}/{r_s}$ is the concentration parameter, and we have also $M(r<\R)=M_{500}=\frac{4\pi}{3}\R^3\Delta_c\rho_c$, where $\Delta_c=500$ and $\rho_c$ is the critical density at the cluster redshift. 

We emphasize that in our analysis we implement two different priors on the mass parameter $M_{500}$; however we also perform the analysis without any restriction on the mass, unlike previous work on other clusters (e.g. Coma cluster in \cite{Terukina:2013eqa}), and therefore we anticipate testing possible degenerate scenarios in the posterior parameter space (allowed at some range of the virial mass), this is discussed at length in the \Cref{App:Mass_prior}.

\subsection{Weak Lensing mass priors}
Chameleon gravity belongs to a subset of scalar-tensor theories for which the gravitational potential inferred by lensing techniques corresponds to the Newtonian potential (i.e. the contribution of fifth force does not affect null geodesics). As such, we can implement information provided by lensing estimation as prior on the ``true" cluster mass $\M$, as done in e.g. \cite{Terukina:2013eqa, Wilcox:2015kna}. We utilize the estimates of $M_{500}$ obtained using weak lensing analyses in \cite{Herbonnet:2019byy}, wherein no information on the shape ($\C$) of the mass profile is available. However, we find that mass information are available only for five clusters in the sample, A85, A1795, A2029, A2142 and ZW1215. 
In \Cref{tab:WLpriors}, we show the mean and $1\sigma$ uncertainties on $M_{500}$ for these clusters, taken from \cite{Herbonnet:2019byy}. We beforehand anticipate that the constraints on mass parameters we shall obtain using the X-COP data will be much tighter than the uncertainty of the weak lensing masses we use as priors.

{\renewcommand{\arraystretch}{1.4}%
    \setlength{\tabcolsep}{6pt}%
    
\begin{table}[!ht]
    \centering
    \caption{We show the weak lensing masses utilized as mass priors for the 5 clusters available from \cite{Herbonnet:2019byy}\footnote{We utilize the mass estimated using the NFW mass profile for consistency, please see the Table A2 in \cite{Herbonnet:2019byy}. See also \Cref{App:Mass_prior} for more comments.}.} 
    \label{tab:WLpriors}
    
    \begin{tabular}{c|c}
    \hline
        
        Cluster & $M_{500}[10^{14}\, M_{\odot}]$\\

        \hline
        \hline
        
        \begin{tabular}{@{}c@{}}A85 \\ \end{tabular} & $5.7 \pm 2.2$  \\
        \begin{tabular}{@{}c@{}}A1795 \\ \end{tabular} & $9.3 \pm 2.2$  \\
        \begin{tabular}{@{}c@{}}A2029 \\ \end{tabular} & $12.1 \pm 2.5$  \\
        \begin{tabular}{@{}c@{}}A2142 \\ \end{tabular} & $9.7 \pm 2.3$   \\
        \begin{tabular}{@{}c@{}}ZW1215 \\ \end{tabular} & $3.5 \pm 2.2$  \\

        \hline

    \end{tabular}
    \end{table}
}

We perform a full Bayesian analysis utilizing \Cref{enq:20,enq:21} to define the likelihood, through the publicly available \texttt{emcee}\footnote{\href{http://dfm.io/emcee/current/}{http://dfm.io/emcee/current/}} package \citep{Foreman-Mackey13, Hogg:2017akh}, which implements an affine-invariant ensemble sampler. To analyze the MCMC chains we utilize either the \texttt{corner} and/or \texttt{GetDist} \footnote{\href{https://getdist.readthedocs.io/}{https://getdist.readthedocs.io/}} \cite{Lewis:2019xzd} packages. Also, we impose uniform flat priors on all the parameters, specifically for the modified gravity parameters $\{\btwo, \phitwo\} \in [0.001, 1.0]$. As the current analysis provides posteriors of exclusion within the parameter space, always including the GR scenario, to the first order we refrain from performing any model selection, which is bound to select GR with higher preference. 

Finally, we implement a simple importance sampling-like routine to combine the constraints in the $\Theta_{\rm MG}$ parameter space, obtained using the individual clusters. Given that the parameters $\Theta_{\rm NFW}$ and $\Theta_{n_e}$ are cluster specific and are not expected to affect the joint constrains on the $\Theta_{\rm MG}$ parameters which are of a global theory. Therefore, we combine the MCMC samples of of the $\Theta_{\rm MG}$ parameters obtained form each of the clusters where the sample density represents the values of the posterior (Bayesian confidence levels). We take a sub-sample of thinned MCMC samples of equal size and re-sample the joint posteriors. Essentially, this approach is equivalent to marginalizing on all the cluster specific parameters, while not being able to see the effect of the joint analysis on them. The results of the combined analysis are given in \Cref{sec:joint}.

\section{Results}
\label{sec:Results}

\begin{figure*}
    \centering
    \includegraphics[scale=0.5]{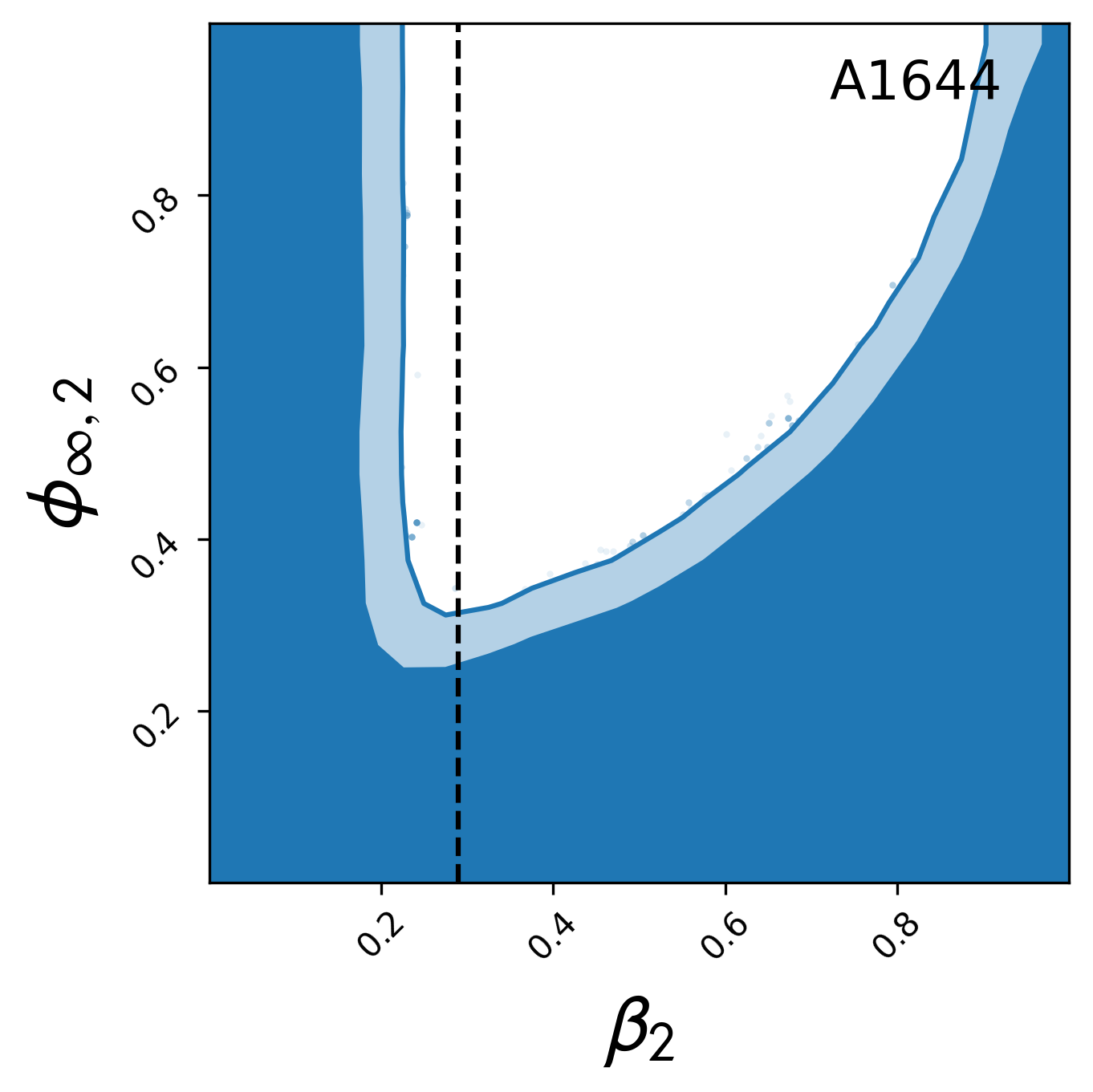}
    \includegraphics[scale=0.5]{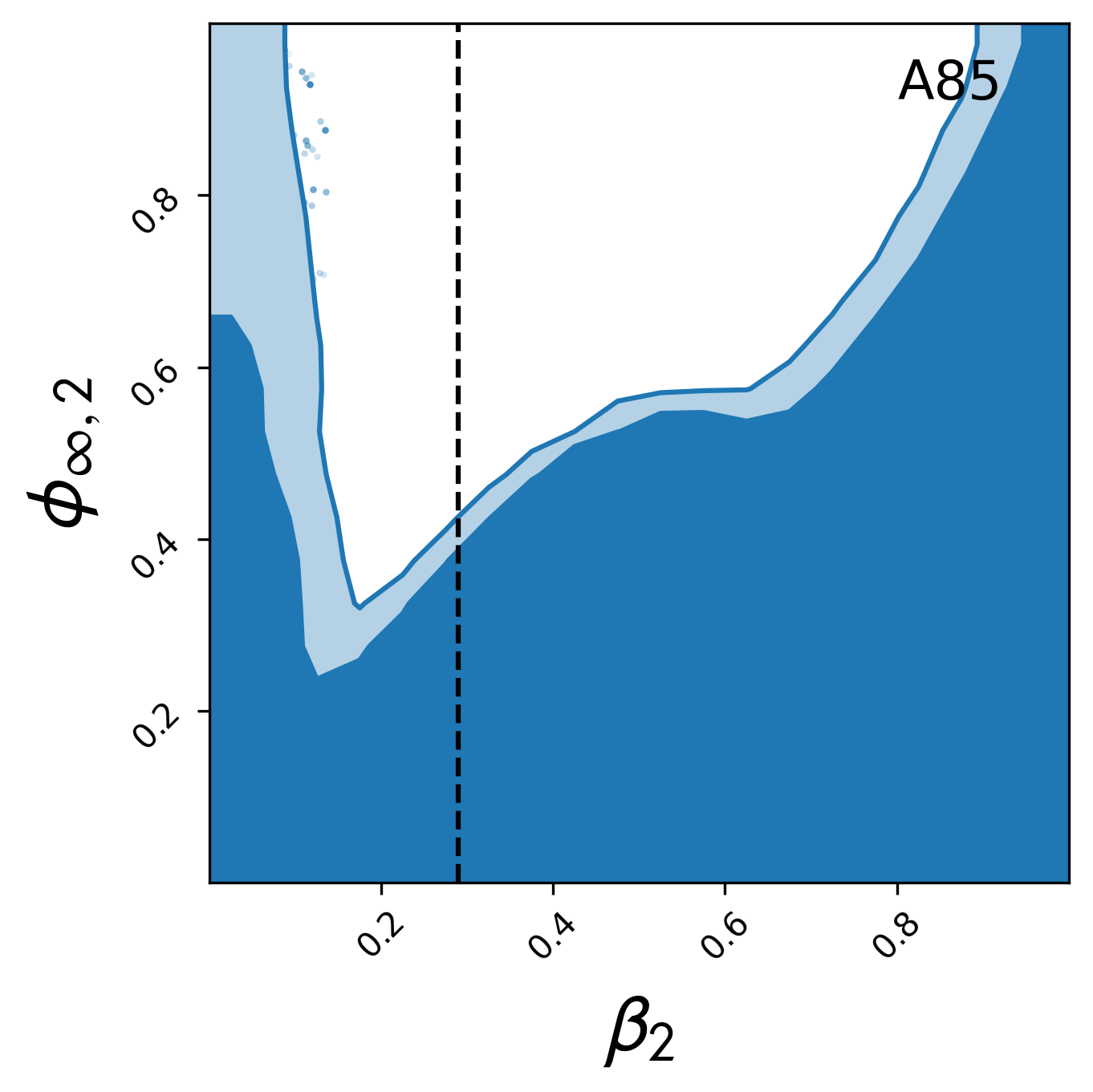}
    \includegraphics[scale=0.5]{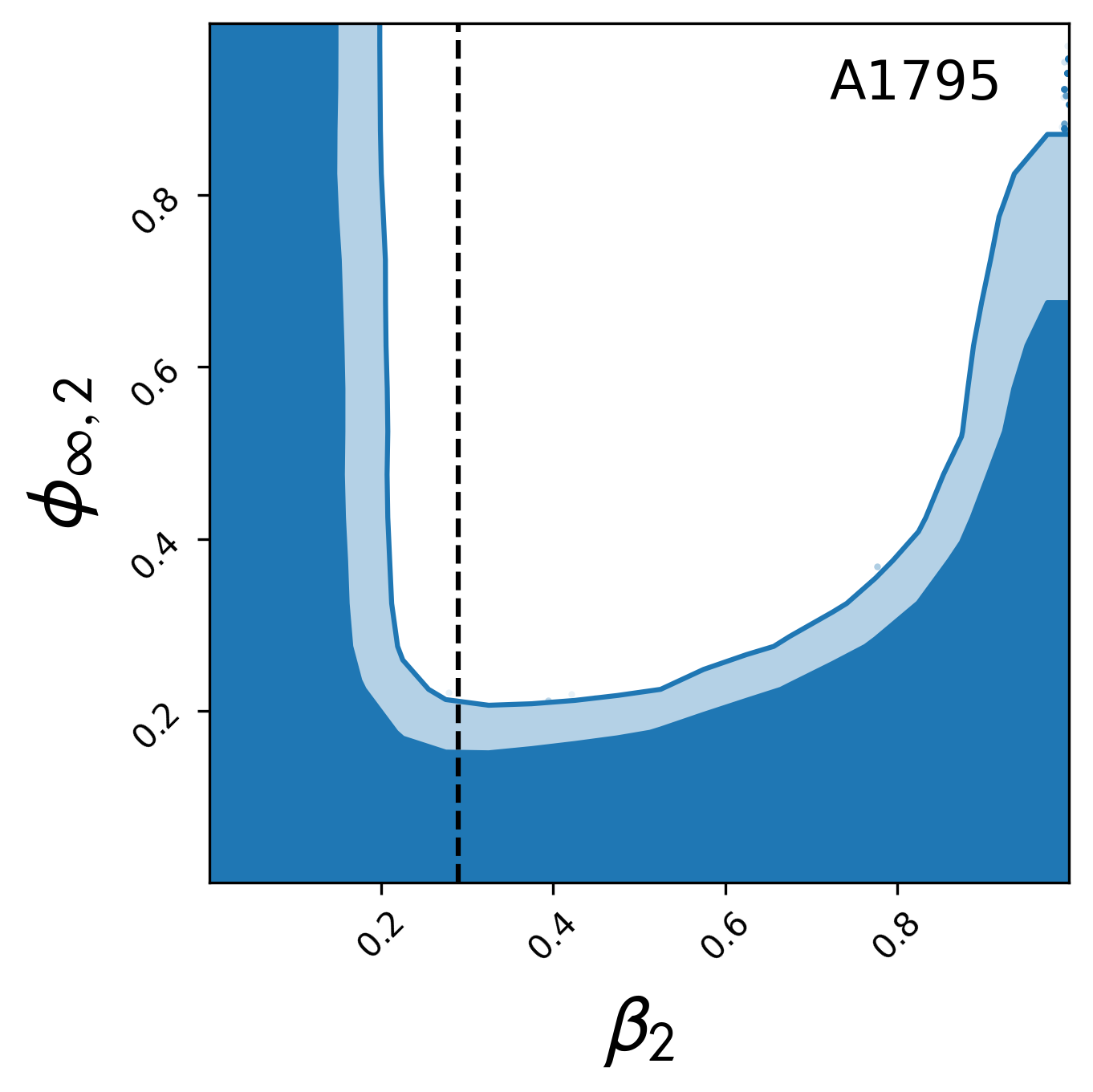}
    \includegraphics[scale=0.5]{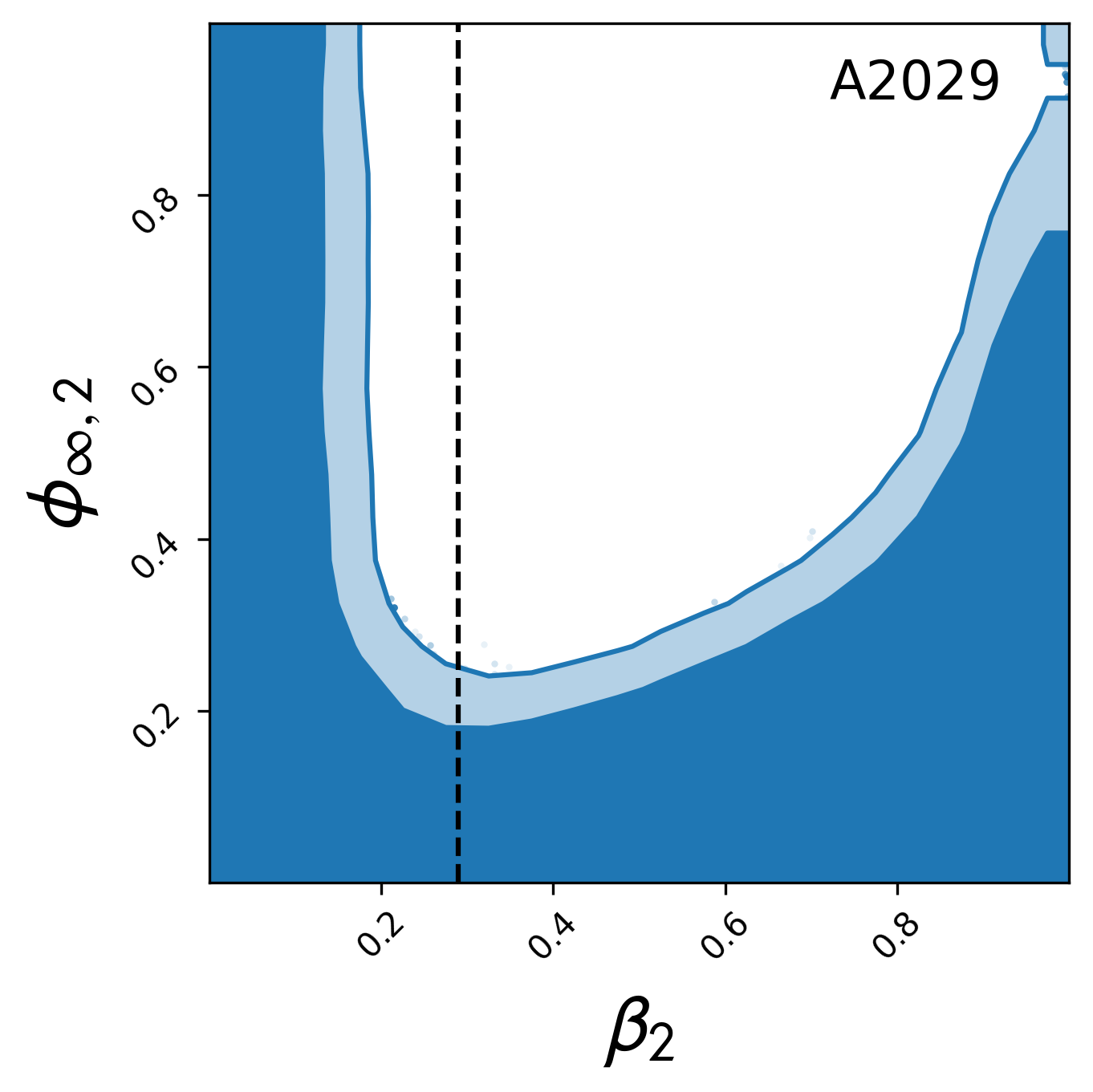}
    \includegraphics[scale=0.5]{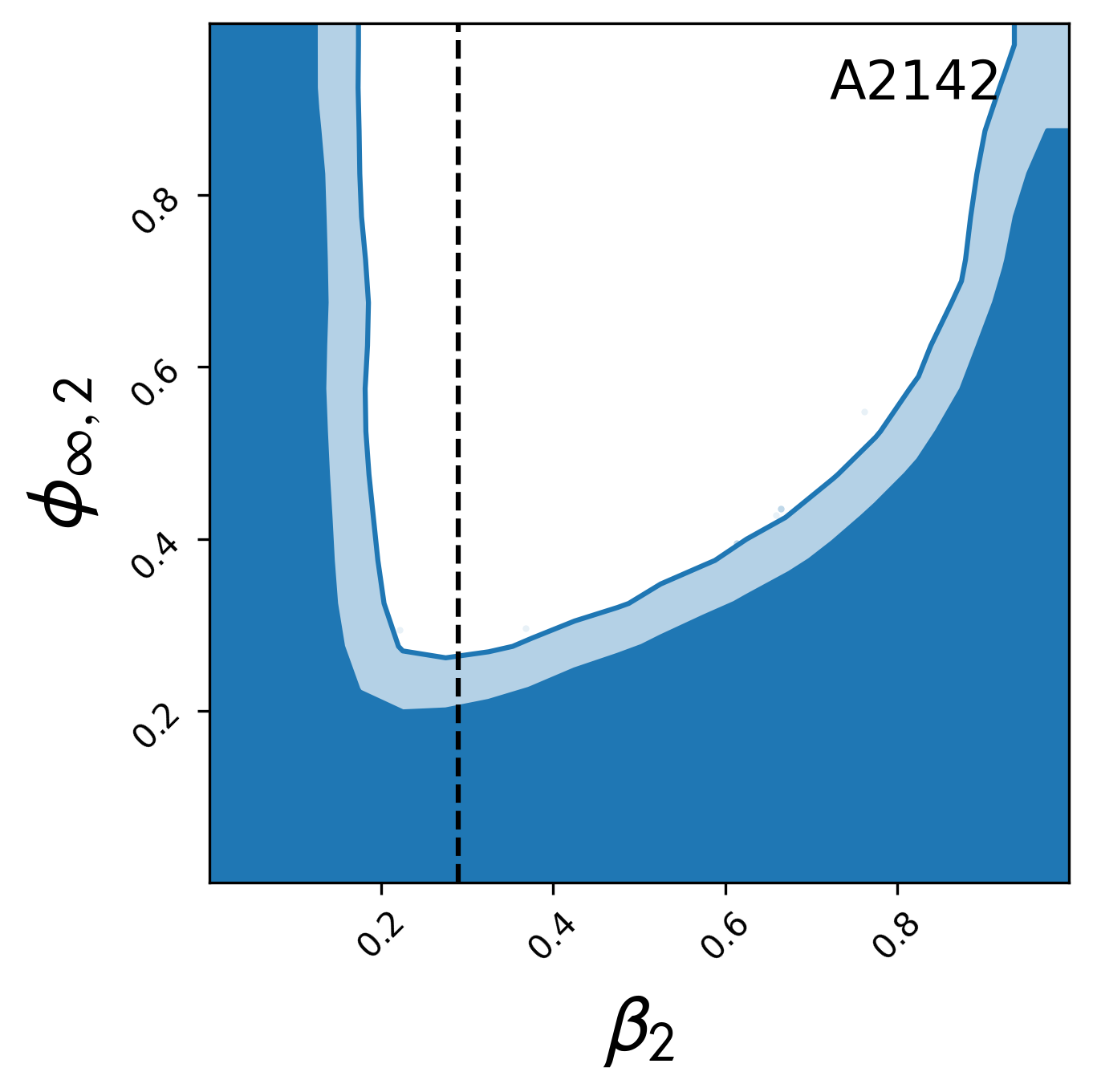}
    \includegraphics[scale=0.5]{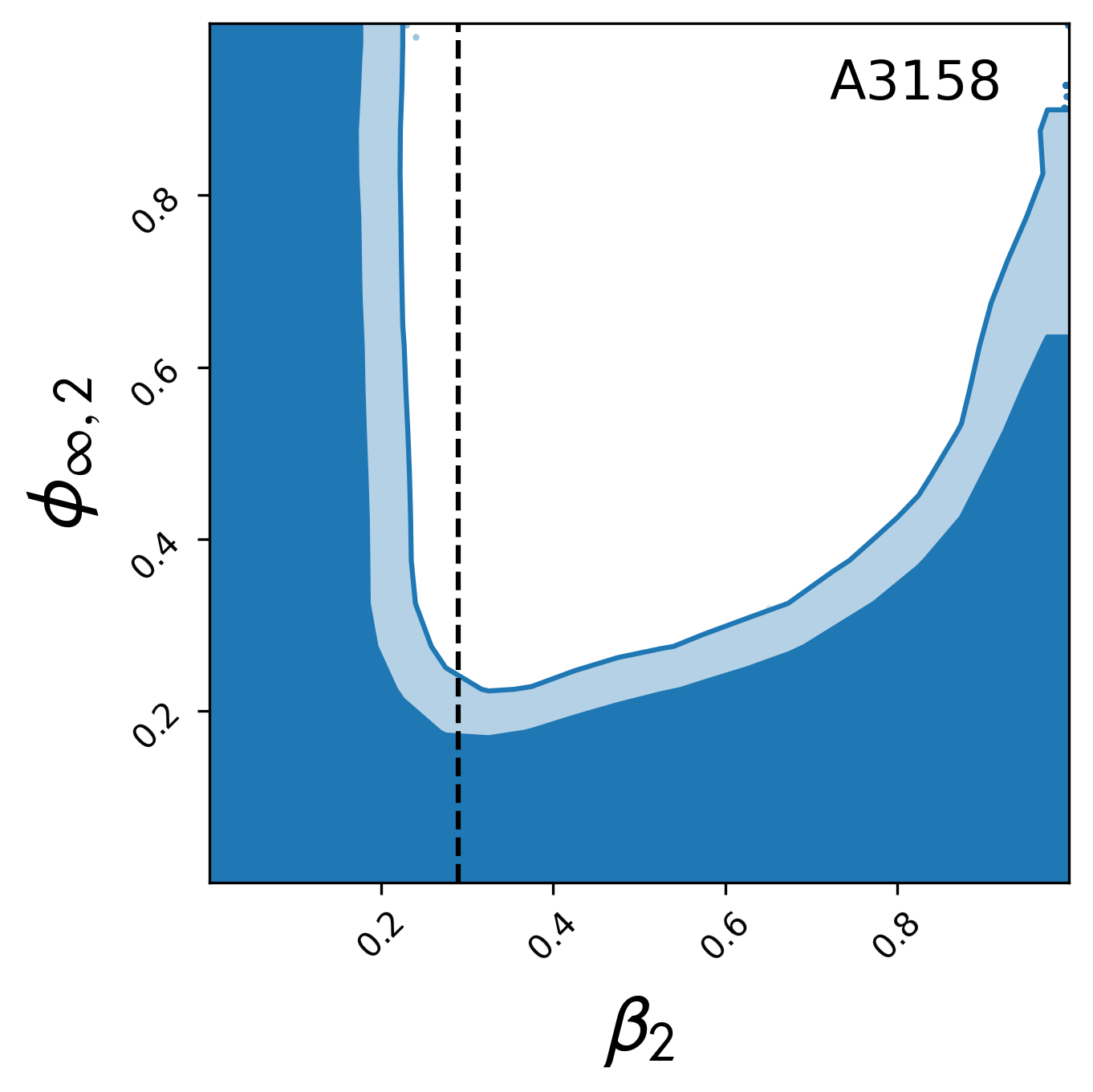}
    \includegraphics[scale=0.5]{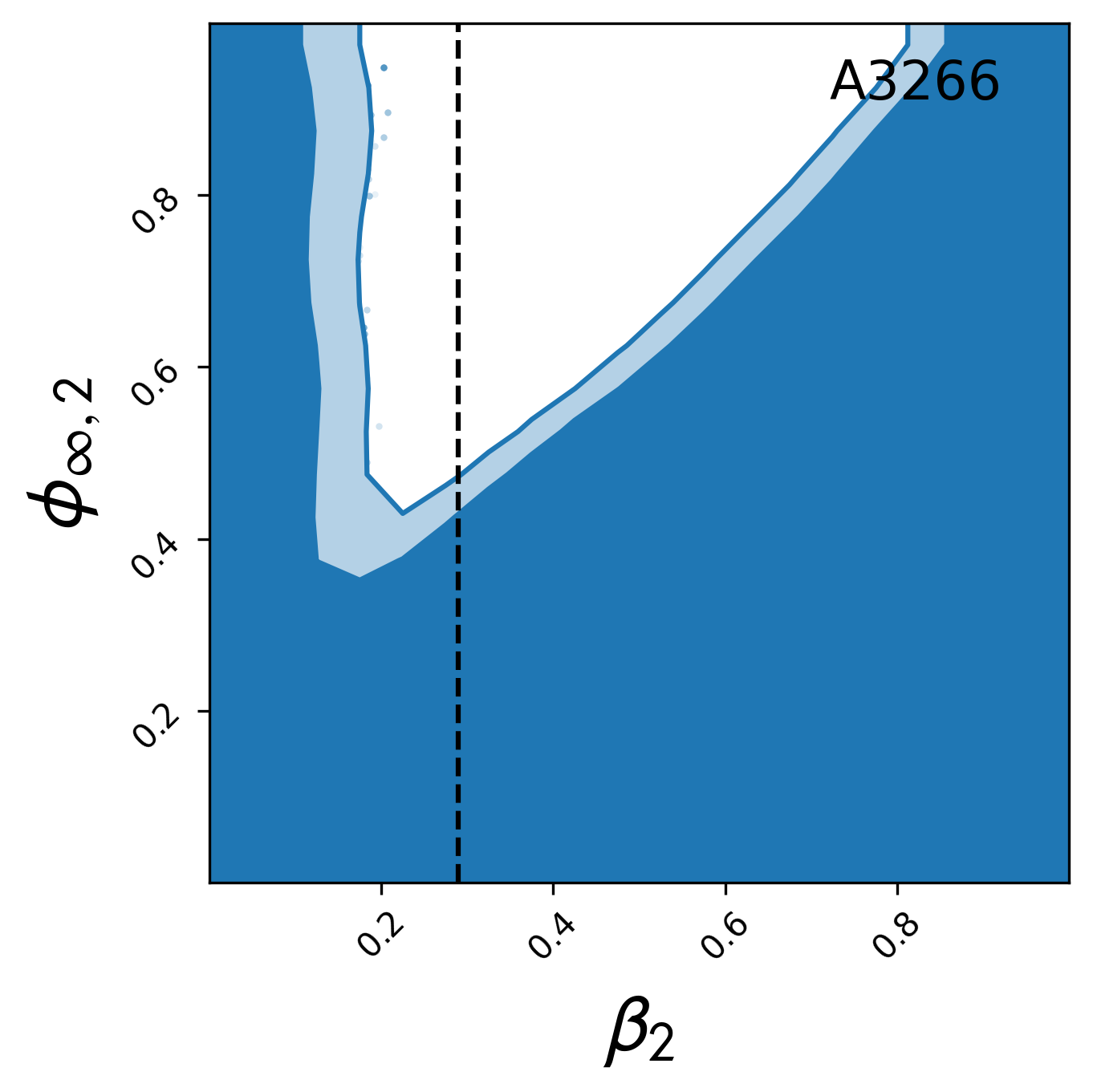}
    \includegraphics[scale=0.5]{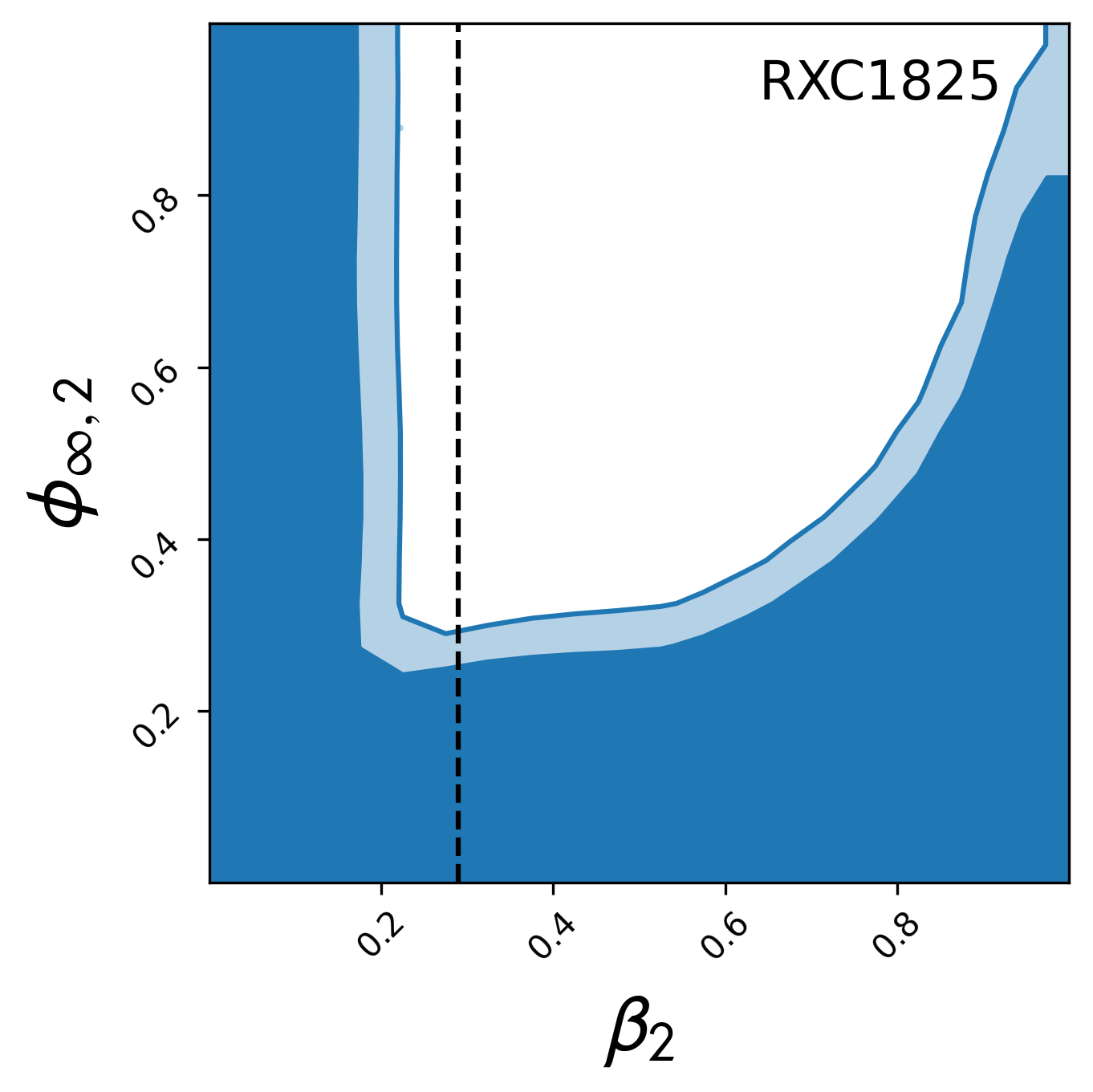}
    \includegraphics[scale=0.5]{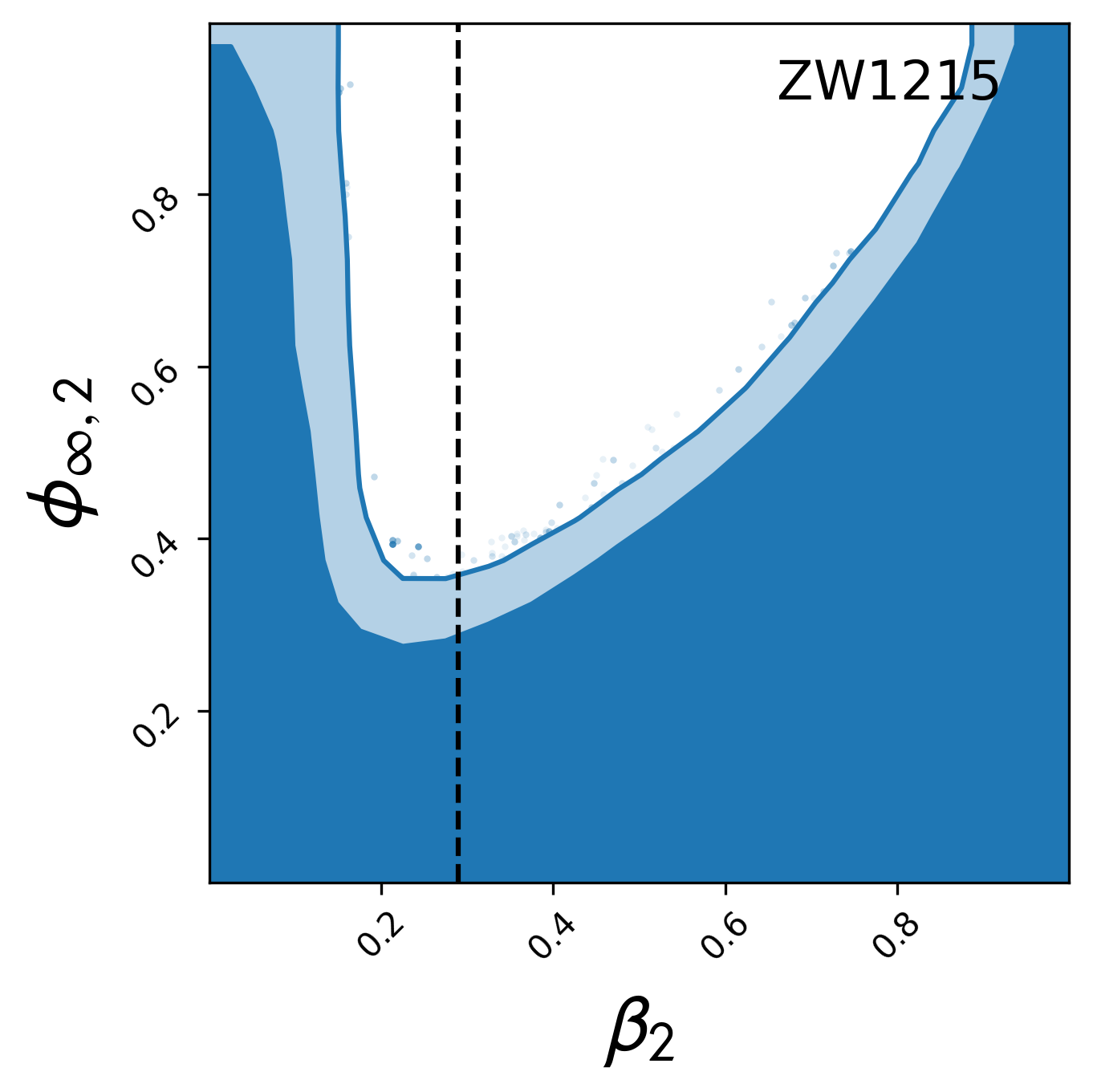}

    \caption{$95\%, 99\%$ C.L. contours for all the clusters utilizing the internal mass prior. The vertical line shows the $\btwo \sim 0.29$, corresponding to the \fofr scenario where $\beta = \sqrt{1/6}$. }
    \label{fig:hydrostartic}
\end{figure*}

We begin by presenting the constraints on the $\{\phi_{\infty,2},\, \beta_2\}$ parameter space for each  of the nine clusters as shown in \Cref{fig:hydrostartic}, utilizing the \textit{internal mass prior}, elaborated in the next paragraph. The blue and light blue regions depicts the allowed parameter space at $2\sigma$ and $3\sigma$, respectively, while the white region consequently is excluded by the current data at $3\sigma$ confidence level. We can already notice for all clusters that at low $\beta_2$ (equally $\beta$), $\phitwo$ is unconstrained: as the coupling constant becomes negligible, the Chameleon field is decoupled from matter and can no longer be constrained. 
Meanwhile, at large values of $\beta$, that is when $\beta_2 \simeq 1$, the coupling is too strong that the entirety of the clusters will be screened, i.e, the screening radius is larger than the size of the cluster in which case also all values of $\phi_{\infty,2}$ are allowed. We also find that at low values of $ \beta_2$, a slightly larger part of the parameter space is excluded compared to the results presented in \cite{Terukina:2013eqa} and \cite{Wilcox:2015kna}, which in our results  extends to $\beta_2 \sim 0.2 $. In \Cref{fig:hydrostartic}, this lower limit is what we see as an almost vertical line in the contours that separates the blue allowed region from the white excluded one for lower values of $ \beta_2$. On the other hand, compared to the same previous results, we find that the lowest possible values for $\phi_{\infty,2}$ are also lower, which further reduces  the allowed region providing tighter constraints in our analysis. This is mainly due to the effect of the internal mass prior, as we will discuss below.

We also point out that in the plots an exponential-shaped bound appears in all of the posteriors of $(\phi_{\infty,2},\beta)$, which is due to the fact that the formalism inherently takes into account the assumption that the critical radius $r_c$ is a positive quantity. From \Cref{eqn:8} it can be shown that this is equivalent to regions below the curve of the following equation, 
\begin{equation}
\label{eqn:22}
    \phi_{\infty,2}=1-\exp \left(-\frac{\beta_2}{1-\beta_2}\frac{\rho_s r_s^2}{10^{-4} M^2_\text{Pl}}\right),\,
\end{equation}

As mentioned before, the contours of \Cref{fig:hydrostartic} are obtained by adding a prior on the parameter $M_{500}$. This is because utilizing only the hydrostatic equilibrium data leads to a strong degeneracy in the $\{\M, \btwo\}$ parameter space, which prevents to place any stringent bounds in most of the cases. In earlier analyses this degeneracy was broken by aiding the hydrostatic data with the mass priors obtained from weak lensing analyses. We further elaborate on this in the \Cref{App:Mass_prior} (c.f. \Cref{fig:compare_3}). 

To assess the constraints while excluding this degeneracy we eliminate the lower mass regions by considering a lower limit of $\beta_2>0.5$ and constrain the posteriors for the $\{\M,\C \}$, following which we construct the mass and concentration priors, also taking into account the corresponding covariance and re-perform the analysis by expanding the range of $\btwo \in (0.0,1.0)$, as shown in \Cref{fig:hydrostartic}. Hereon we denote this prior as \textit{internal mass prior} and elaborate in \Cref{sec:deg}. We find that this degeneracy is usually present within $\beta_2<0.5$, corresponding to $\beta <1$, accounting for a decrease in the values of $\M$ while the values of $\beta$ increase, following the expression of the thermal pressure in \Cref{eqn:15}. In clusters A85 and RXC1825 however, we find this degeneracy to extend beyond $\beta>1$. In particular for A85, we see that the internal mass prior is completely unable to even reduce the degenerate region.

We then show quantitative results of our analysis in \Cref{tab:Constraints} we show the results of our analysis for the nine X-COP clusters used in this work. We present in the first column the $68\%$ C.L. of the concentration parameter $c_{500}$ and the mass $M_{500}$ with the internal mass prior elaborated above. We also present the $95\%$ C.L. limits on the value of the field $\phi_{\infty,2}$ for $\beta=\sqrt{1/6}$, which corresponds to the $f(R)$ sub-class of  Chameleon model, presented in \Cref{Sec:Modell}. In the subsequent columns we present the values at $95\%$ C.L.  we obtain for the field $\phi_{\infty,2}$ when imposing the weak lensing mass prior presented in \Cref{fig:MPRNE} and no mass prior, respectively, which we added for completeness. Within parentheses we show the conversion of $\phi_{\infty,2}$ into $|f_{R0}|$ to get explicit constraints on $f(R)$ models. As can be seen comparing the internal mass prior and no prior scenario, the constraints deteriorate substantially for all the clusters except A85 and A3158. In the case of the case of A85 this posteriors are dominated by the presence of the degeneracy in $\{\M, \btwo\}$ parameter space. On the other hand, the cluster A3158 shows least observed degeneracy. As for mass profile constraints, $c_{500}$ and $M_{500}$, presented in the first two columns of \Cref{tab:Constraints}, are the same for as the GR values  up to a $1\sigma $ confidence level \cite{Haridasu:2021hzq},they are very much in agreement with those estimated for DHOST gravity as presented therein.

One can also notice that the case where we consider an internal mass prior the constrains get considerably tighter, for instance the A1795 field value is eight times tighter than the one with no mass prior and three times than the one with the weak lensing prior (which is yet a good constraint compared to the one with no mass prior).
Also, we point out that the two dimensional posteriors are visually  much tighter than those previously presented for Coma Cluster \cite{Terukina:2013eqa} and XMM Clusters in  \cite{Wilcox:2015kna}. We later perform a more qualitative comparison for the  $|f_{R0}|$, in the $\fofr$ scenario.

\begin{figure}
    \centering
    \includegraphics[scale=0.6]{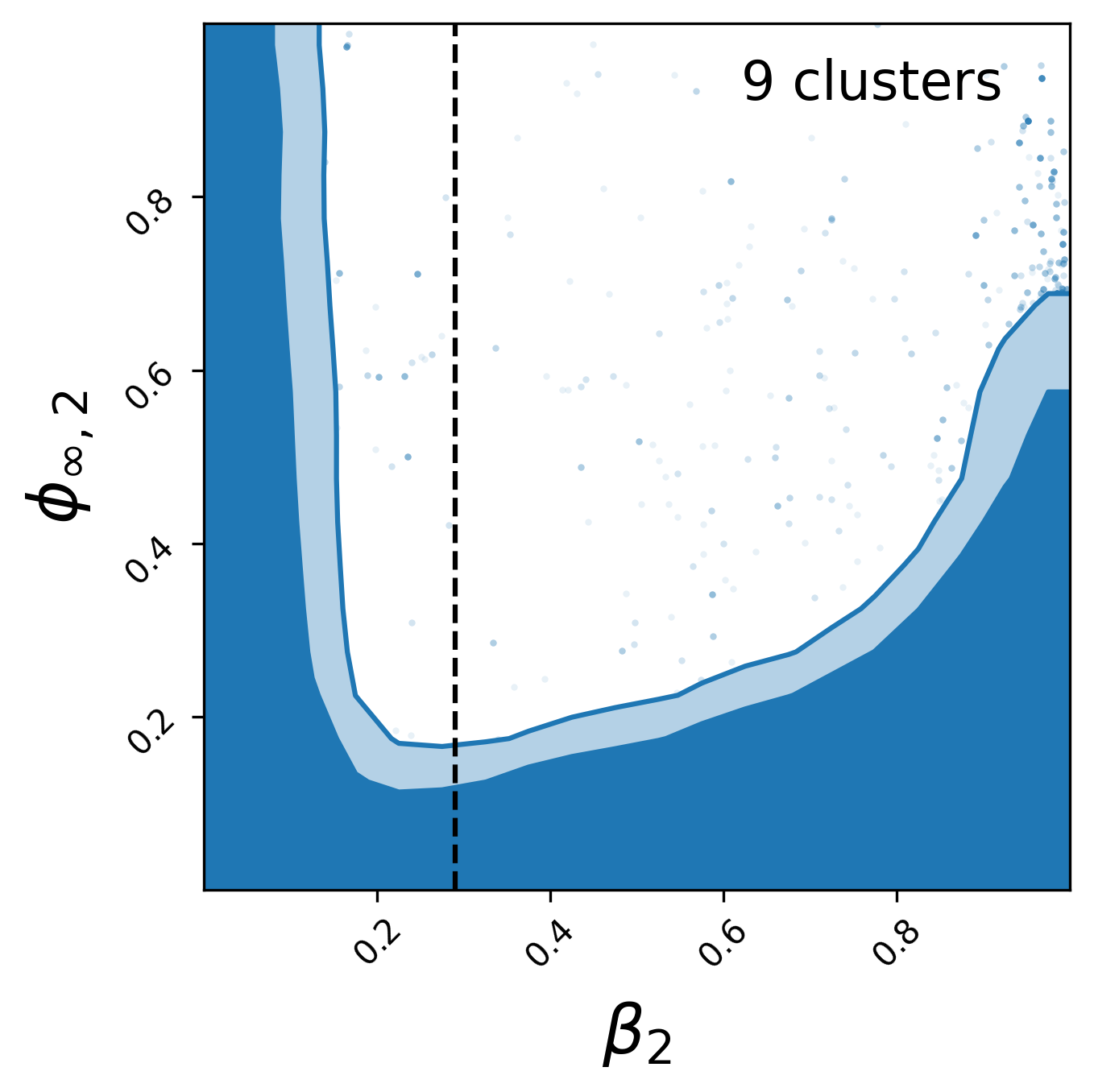}
    \caption{95$\%$ and 99$\%$  C.L. contours for joint constraint utilizing the internal mass priors for 9 of the clusters. The dashed vertical line represents the $\beta = \sqrt{1/6}$, corresponding to the \fofr scenario. }
    \label{fig:BPR_Joint}
\end{figure}

{\renewcommand{\arraystretch}{1.8}%
    \setlength{\tabcolsep}{8pt}%

\begin{table*}[!ht]
    \centering
    \caption{Constraints on parameters $\{c_{500},\,M_{500}\}$ and $\phi_{\infty,2}$ at $68\%$ and $95 \%$ C.L., respectively, from the analysis of each cluster in the sample. The limits are obtained with 3 different methods: from column two to four, inclusion of an internal mass prior to avoid the statistical degeneracy between mass and MG parameters, as discussed in Appendix \ref{App:Mass_prior}. Column five: WL mass prior - which is available only for five clusters. Column six: no mass prior is taken into account. Note that the bounds on $M_{500}$ and $c_{500}$ are shown only for the first case. The constraints on $\phi_{\infty,2}$ are all presented for $\beta=\sqrt{1/6}$ which corresponds to $f(R)$ gravity. The corresponding bounds on the parameter $|f_{R0}|$ are shown inside the parentheses.} 
    \label{tab:Constraints}
    
    \begin{tabular}{ccc|ccc}
    \hline
        
        \multicolumn{1}{c|}{Cluster} & \multicolumn{3}{c|}{Internal mass prior} &  \multicolumn{1}{c|}{WL mass prior} & No mass prior  \\
        \hline
         & $c_{500}$ & $M_{500}\,$ &  \multicolumn{3}{c}{$\phi_{\infty,2}(|f_{R0}|\, [10^{-5}])\,$} \\
        
         & & $ [10^{14}\, M_{\odot}]$ & & &  \\
        \hline
        \hline
        
        \begin{tabular}{@{}c@{}}A85 \\ \end{tabular} & $2.05_{-0.07}^{+0.07}$ & $6.13_{-0.18}^{+0.18}$ & $ 0.272\,(2.592)$ & $0.279(2.671)$ & $0.276(2.637)$  \\
        \begin{tabular}{@{}c@{}}A1644 \\ \end{tabular} & $1.13_{-0.14}^{+0.11}$ & $2.95_{-0.20}^{+0.20}$ & $ 0.226\,(2.092)$ & $/$ & $0.942 (23.25)$   \\
        \begin{tabular}{@{}c@{}}A1795 \\ \end{tabular} & $3.17_{-0.14}^{+0.14}$ & $4.48_{-0.15}^{+0.15}$ & $ 0.146\,(1.289)$ & $0.319(3.137)$ & $0.874(16.91)$   \\
        \begin{tabular}{@{}c@{}}A2029 \\ \end{tabular} & $3.20_{-0.13}^{+0.13}$ & $7.70_{-0.24}^{+0.24}$ & $ 0.208\,(1.904)$ & $0.396(4.117)$ & $0.942(23.25)$    \\
        \begin{tabular}{@{}c@{}}A2142 \\ \end{tabular} & $2.22_{-0.08}^{+0.08}$ & $8.32_{-0.19}^{+0.19}$ & $ 0.198\,(1.802)$ & $0.213(1.956)$ & $0.498(5.627)$    \\
        \begin{tabular}{@{}c@{}}A3158 \\ \end{tabular} & $1.98_{-0.14}^{+0.14}$ & $3.96_{-0.16}^{+0.16}$ & $0.216\,(1.987)$ & $/$ & $0.281(2.694)$  \\
        \begin{tabular}{@{}c@{}}A3266 \\ \end{tabular} & $1.61_{-0.11}^{+0.11}$ & $7.21_{-0.32}^{+0.28}$ & $0.245\,(2.295)$ & $/$ & $0.804(13.30)$   \\
        \begin{tabular}{@{}c@{}}RXC1825 \\ \end{tabular} & $2.54_{-0.24}^{+0.20}$ & $3.90_{-0.15}^{+0.17}$ & $0.146\,(1.289)$ & $/$ & $0.358(3.619)$  \\
        \begin{tabular}{@{}c@{}}ZW1215 \\ \end{tabular} & $1.40_{-0.09}^{+0.09}$ & $7.43_{-0.29}^{+0.29}$ & $ 0.342\,(3.417)$ & $0.892(18.17)$ & $0.567(6.834)$  \\
        \hline 
        Joint & -- & -- & 0.106\,(0.915) & 0.130\,(1.139) & -- \\ 

        \hline

    \end{tabular}
    \end{table*}
}

\subsection{Constraints using weak lensing mass prior}

\begin{figure*}
    \centering
    \includegraphics[scale=0.5]{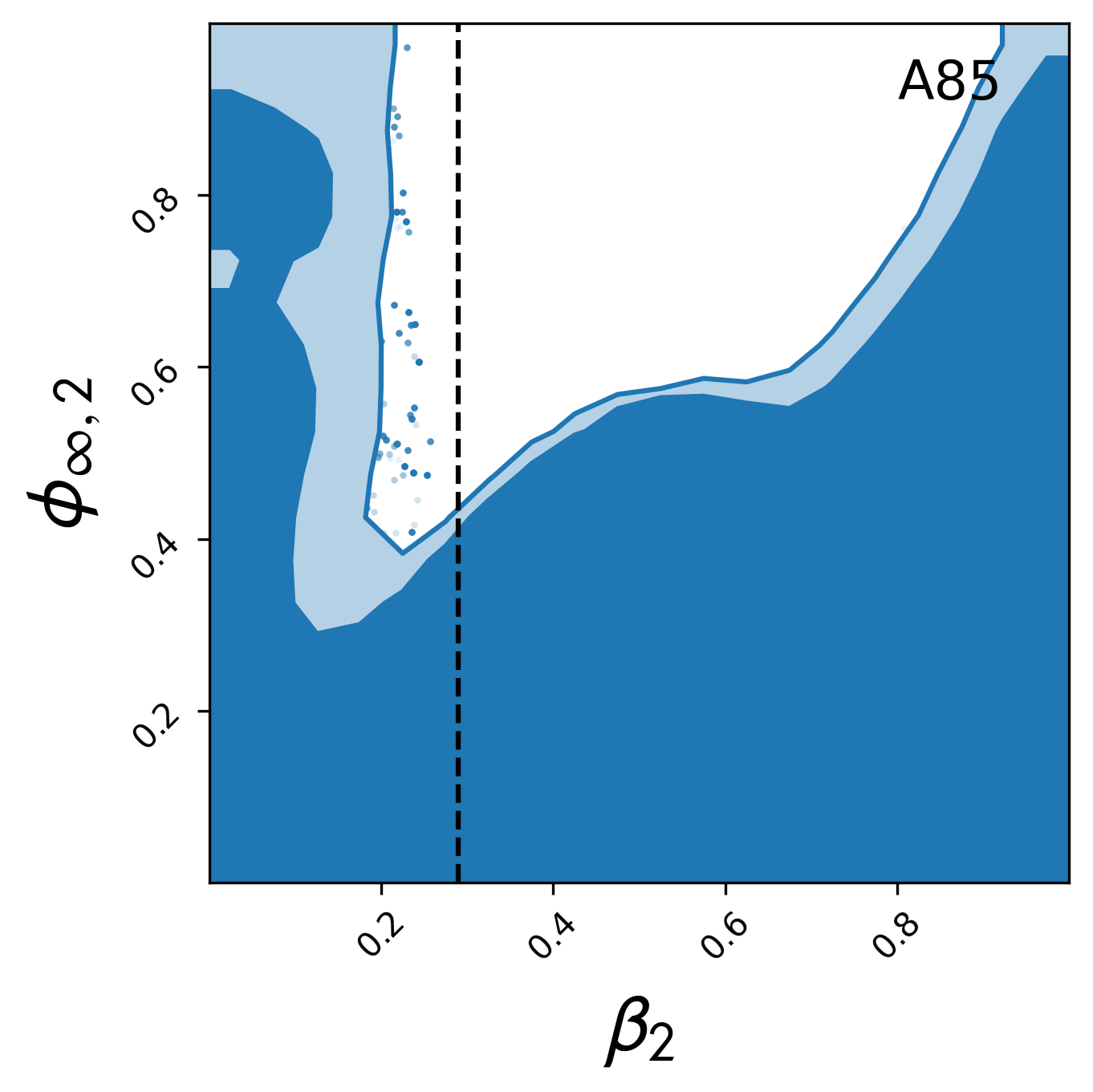}
    \includegraphics[scale=0.5]{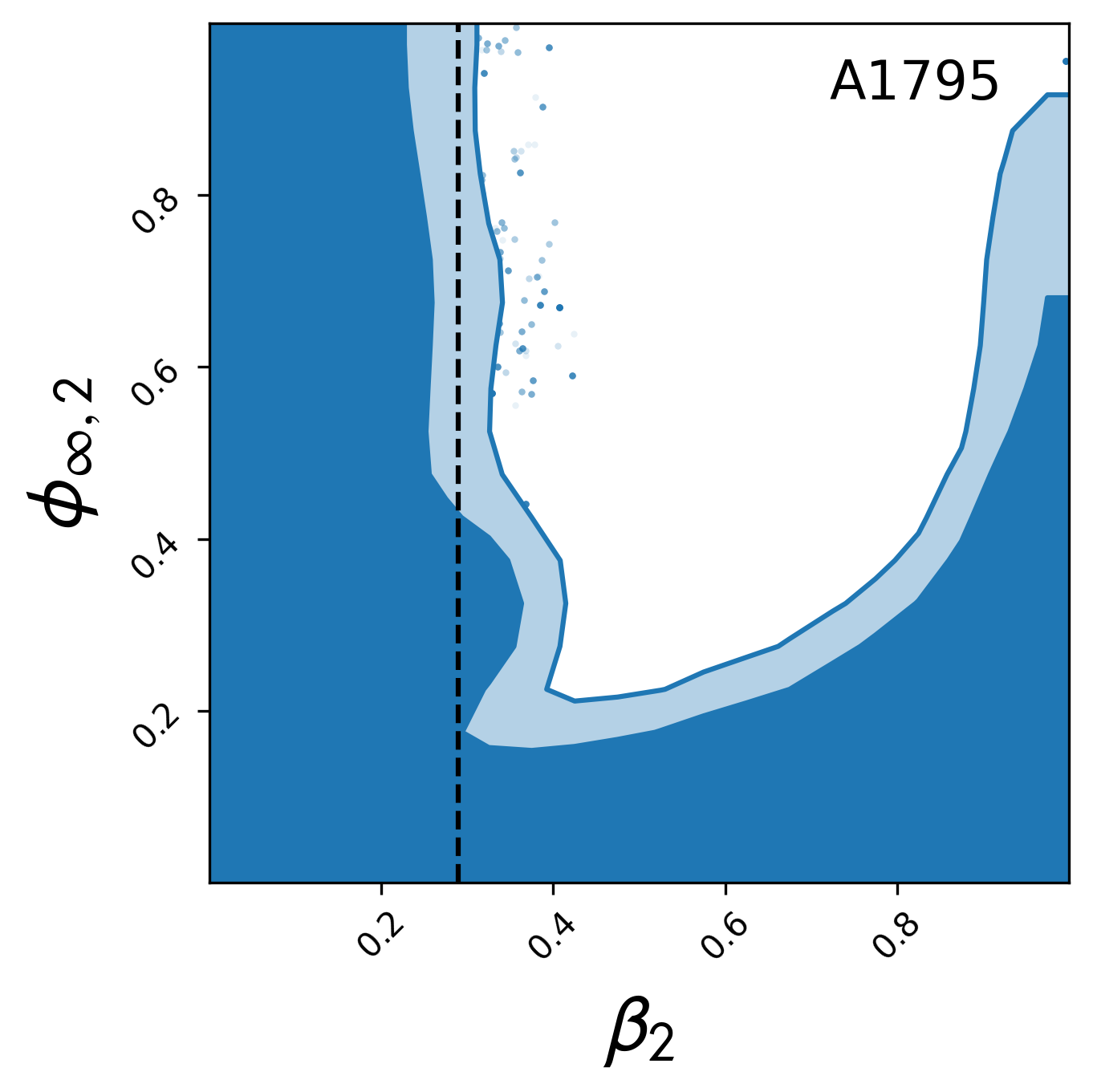}
    \includegraphics[scale=0.5]{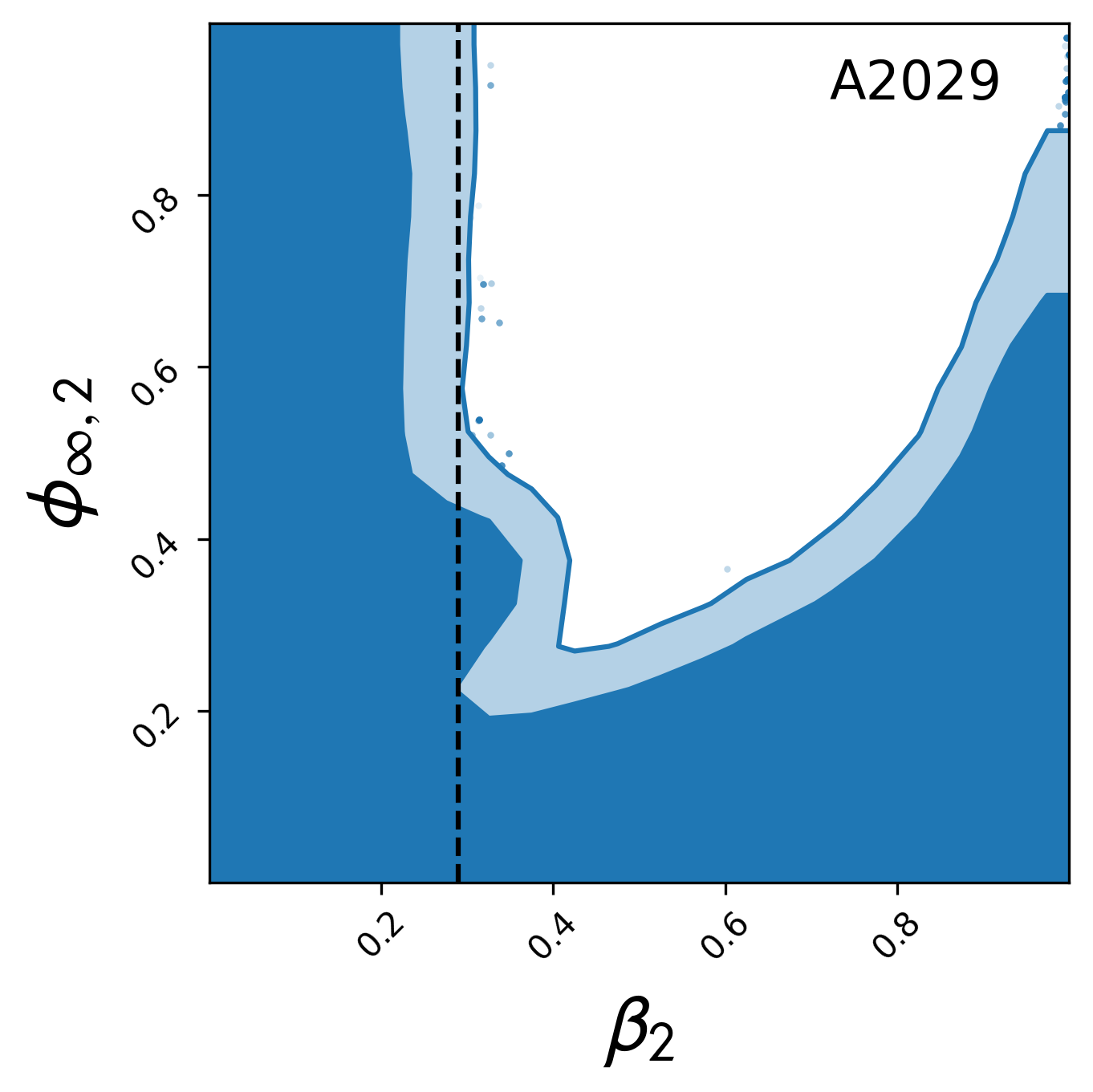}
    \includegraphics[scale=0.5]{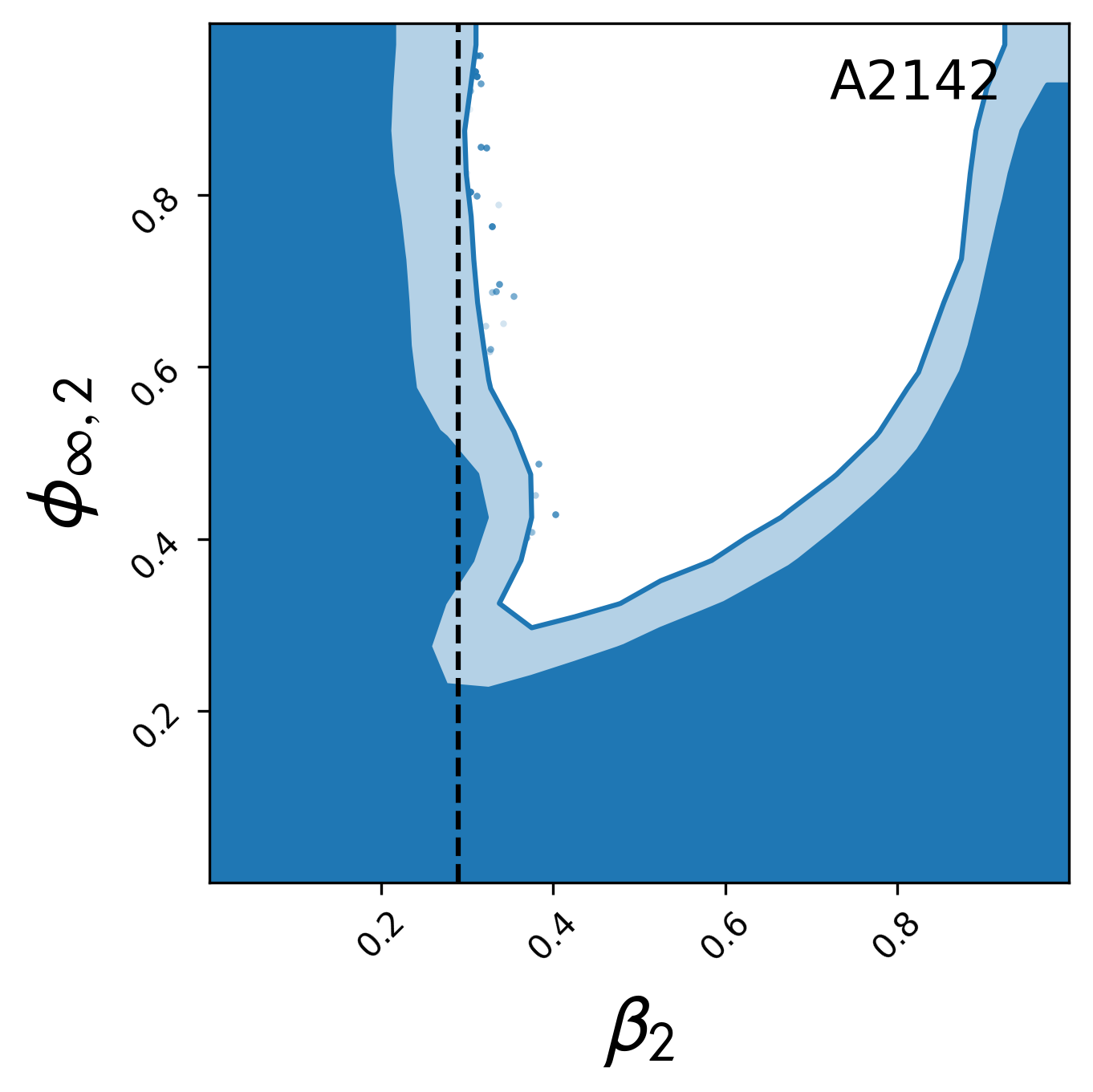}
    \includegraphics[scale=0.5]{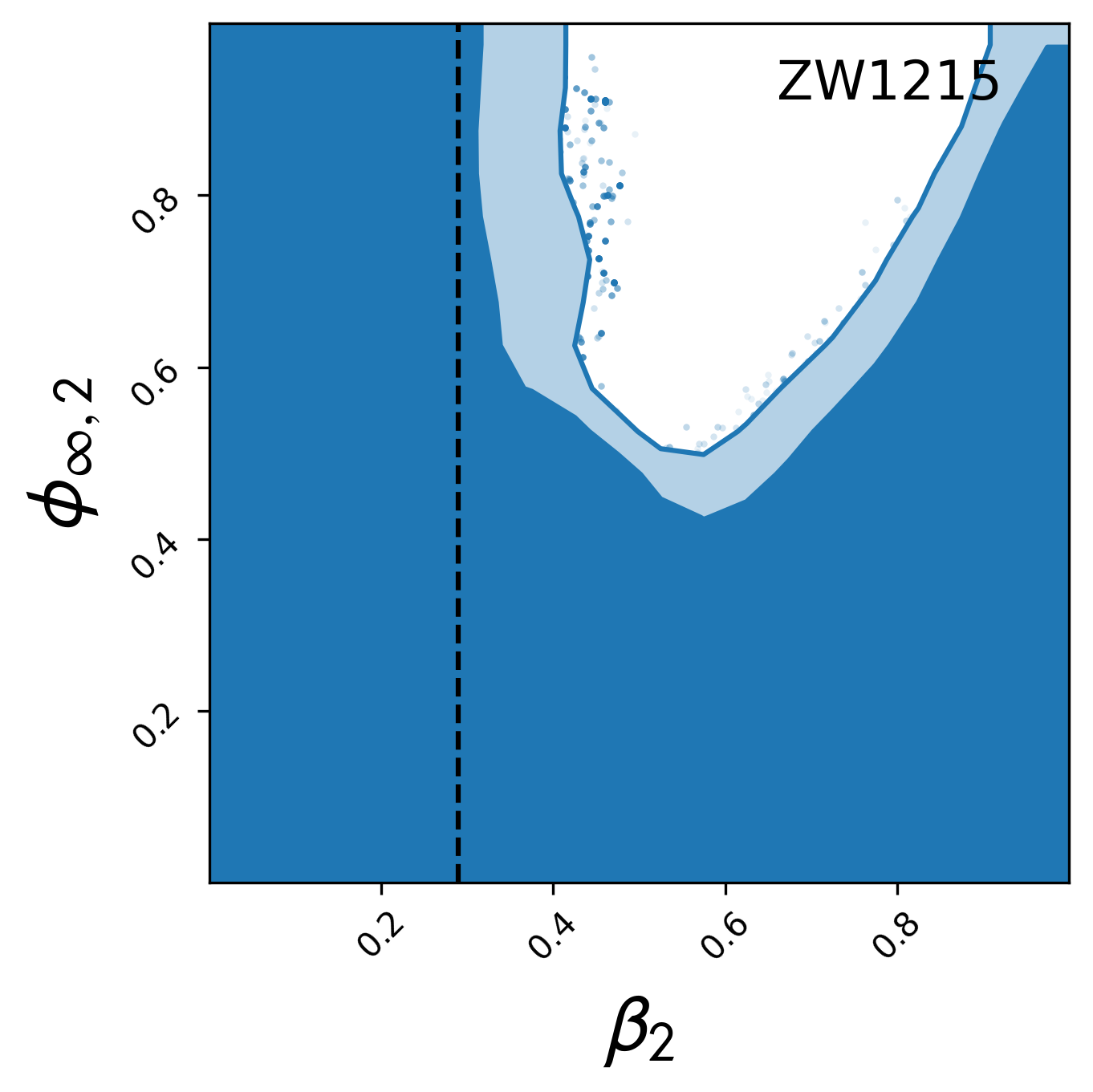}
    \includegraphics[scale=0.5]{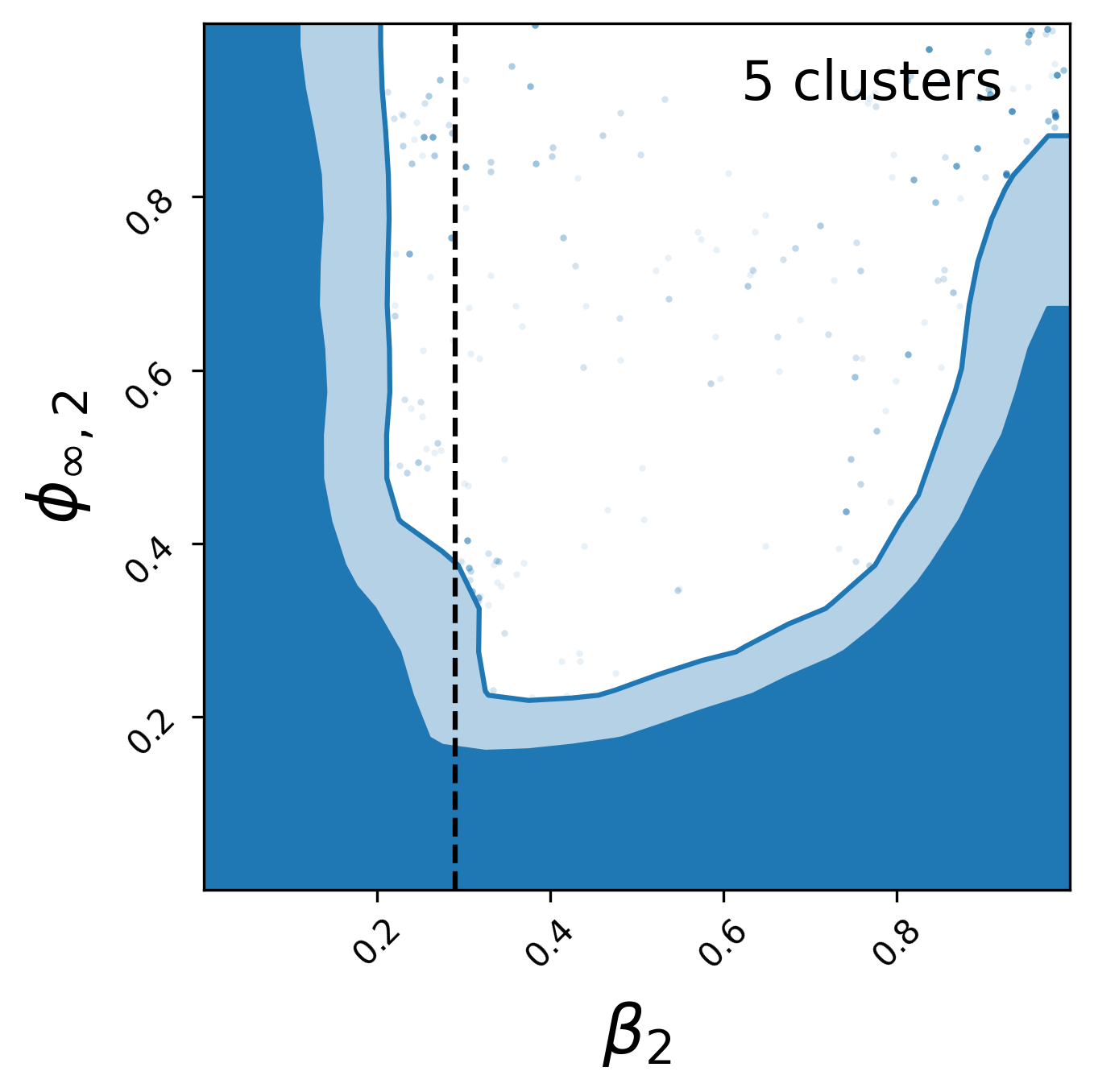}
    
    \caption{95$\%$ and 99$\%$  C.L. contours for weak lensing priors for 5 of the clusters. In the lower right panel we present the joint constraint obtained combining all the five clusters. The dashed vertical line represents the $\beta = \sqrt{1/6}$, corresponding to the \fofr scenario. }
    \label{fig:MPRNE}
\end{figure*}

{In this section we present the constraints obtained on the five clusters for which the weak lensing mass priors are included from the results of \cite{Herbonnet:2019byy}, namely A1795, A2029, A2142, A85 and ZW1215. While aiding the analysis as an independent prior on the mass of the cluster, this also reduces the aforementioned degeneracy between the $\{\btwo, \M\}$ parameters. The constraints on the modified gravity parameters are shown in  \Cref{fig:MPRNE}.} Note that for the cluster ZW1215 alone the inclusion of the WL prior does not aid the constraint and on the other hand, slightly deteriorates the upper limits. This is clearly the case, as the prior itself is an estimated lower value aiding to the degeneracy region, with a mass of order $3.5\, [10^{14}\, M_{\odot}] $ \footnote{Note that \cite{Herbonnet:2019byy} also present the weak lensing mass for the ZW1215 cluster, including others using varied methods, which is higher $\sim 7 \times 10^{14} [M_{\odot}]$.}. However, this does not hinder our ability to constraint the modified gravity parameters in the joint analysis, as discussed in \Cref{sec:joint}. And it is apparent that the degeneracy that remains in the A85 cluster does not affect the joint constraint being guided by the other cluster. 

As expected, we notice that the WL mass prior is capable of reducing the degeneracy elaborated earlier and make the posteriors in the $\{\phitwo,\btwo\}$ slightly more constrained. Note however that the WL mass estimates do present a mass bias (b = $\M^{\rm WL}/\M^{\rm HS}$) which is slightly larger than unity ($b\sim 1.18\pm 0.12$) \cite{Ettori:2018tus} at $R_{500}$. However, in terms of the constraints, even the inclusion of the WL mass prior is not able to remove the  degeneracy completely, which can be seen as  mild bump in the posteriors presented in \Cref{fig:MPRNE}. This is clearly due to the larger uncertainties on the WL masses in comparison to the constraints on $\M$ obtained from the hydrostatic equilibrium. Our formalism here validates that having a well-constrained independent mass estimate from WL method, where in the weak lensing potential is unaffected by the chameleon gravity can be very beneficial for constraining the parameters. 

\subsection{Parameter degeneracy}
\label{sec:deg}
Alongside obtaining the constraints on $\{\phitwo, \btwo\}$ parameters, we also comment on the  degeneracy(s) that we notice between the cluster mass profile parameters and the modified gravity parameters. As can be seen in  \Cref{fig:compare_3}, an increase in $\beta_2$ or $\phi_{\infty,2}$ is compensated by lower values of  $M_{500}$. This is not surprising, given the structure of the modified Newtonian potential in \Cref{eqn:15}.   This degeneracy is also visible in the marginalized posterior distribution of $\phi_{\infty,2}$ and $\beta_2$ as a bump, which emphasize the necessity of a mass prior along with hydrostatic equilibrium data. 
Indeed, we can notice that the degeneracy reduces as soon as we add additional information on $M_{500}$, and the tighter this mass prior is the less degeneracy we have.  Earlier hydrostatic equilibrium analysis which always considered the WL counterpart in did not find such a  degeneracy, for instance using COMA cluster in \cite{Terukina:2013eqa} and XMM  cluster in \cite{Wilcox:2015kna}.

We can also see this quantitatively from the condition we impose in our model to estimate the \textit{screening} radius, which gives a direct relation between $\M$ and $\beta$. In particular, replacing \Cref{eqn:18,eqn:19} into \Cref{eqn:8} one can write, 
\begin{equation}
\label{eqn:23}
    1+\frac{r_c}{r_s}\sim \frac{1}{M_{500}^{-5/2}}\frac{\beta}{\phi_\infty} f(c_{500})
\end{equation}
Here $f(c_{500})$ a function that only depends on the shape of the profile ($c_{500}$).
At this stage, if we impose the condition that maps all negative $r_c$ to $r_c\sim 0$ we get from above $\frac{\beta}{\phi_\infty} \sim M_{500}^{-5/2}$, this means that when the coupling constant $\beta$ is low, the mass gets higher, which creates a region where the higher the mass, the lower the coupling and vice versa, as can be seen in \Cref{fig:compare_3}. Also within the hydrostatic equilibrium equation, the contribution of the gravitational force and the fifth force, are scaled by $\M$ and $\beta$, respectively. The summation of these two forces provide the derivative of the pressure and not knowing the integration constant $P(r = 0)$ beforehand allows only the shape to be constrained and hence the degeneracy between these two forces is propagated to the corresponding parameters. 

One can also notice in the $\{\phi_{\infty,2},M_{500}\}$ plot of \Cref{fig:compare_3} that the same degeneracy holds: lower values of the mass correspond to slightly higher $\phi_{\infty,2}$ (equally $\phi_{\infty}$ ). 
This region appears only for low mass values and coupling constant $\btwo<0.5$ (i.e, $\beta<1.0$). As for the higher masses limit, this degeneracy disappears with the coupling strength approaching $\btwo \to 0 $. Therefore to avoid such a statistical degeneracy we construct an internal mass prior based on the mass values we get for $\beta_2>0.5$ and then run the MCMC chain again to get the new posteriors, and this will erase the degeneracy issue as shown in \Cref{fig:hydrostartic}. Alternatively, adding the WL mass prior will remarkably reduce the degeneracy region as shown in \Cref{fig:compare_3} and the posteriors are shown in \Cref{fig:MPRNE}.

\subsection{Joint analysis}
\label{sec:joint}
Considering that the clusters utilized in the analysis are independent datasets, we explore the possibility to obtain joint constraints on the modified gravity parameters $\{\phitwo, \btwo\}$. {In principle, the background field should evolve in cosmic time. However, given the small redshift range ($0.04\le z \le0.09$) spanned by the sample, we can safely neglect any redshift dependence and assume $\phi_\infty(z)\sim \phi_\infty(z=0)$, essentially constraining the local value of the field.} In \Cref{fig:BPR_Joint} and the lower right panel of the \Cref{fig:MPRNE}, we show the joint constrains using 9 clusters and 5 clusters with the WL mass priors, respectively. Firstly, the overall posterior parameter space in \Cref{fig:BPR_Joint} is greatly reduced when the 9 clusters are combined displaying the ability of the current hydrostatic data to constrain the chameleon screening model, improving the constraints from the earlier analysis in \cite{Terukina:2013eqa, Wilcox:2015kna}. Note that the internal mass prior plays a very important role in allowing such tight constraints. The joint constrains using of the 5 clusters using the WL mass prior as well are tighter constrains with a mild residual of the degenerate region.

\subsection{Joint constraints on $\fofr$ gravity}
\label{sec:joint_fofr}

\begin{figure}
    \centering
    \includegraphics[scale=0.46]{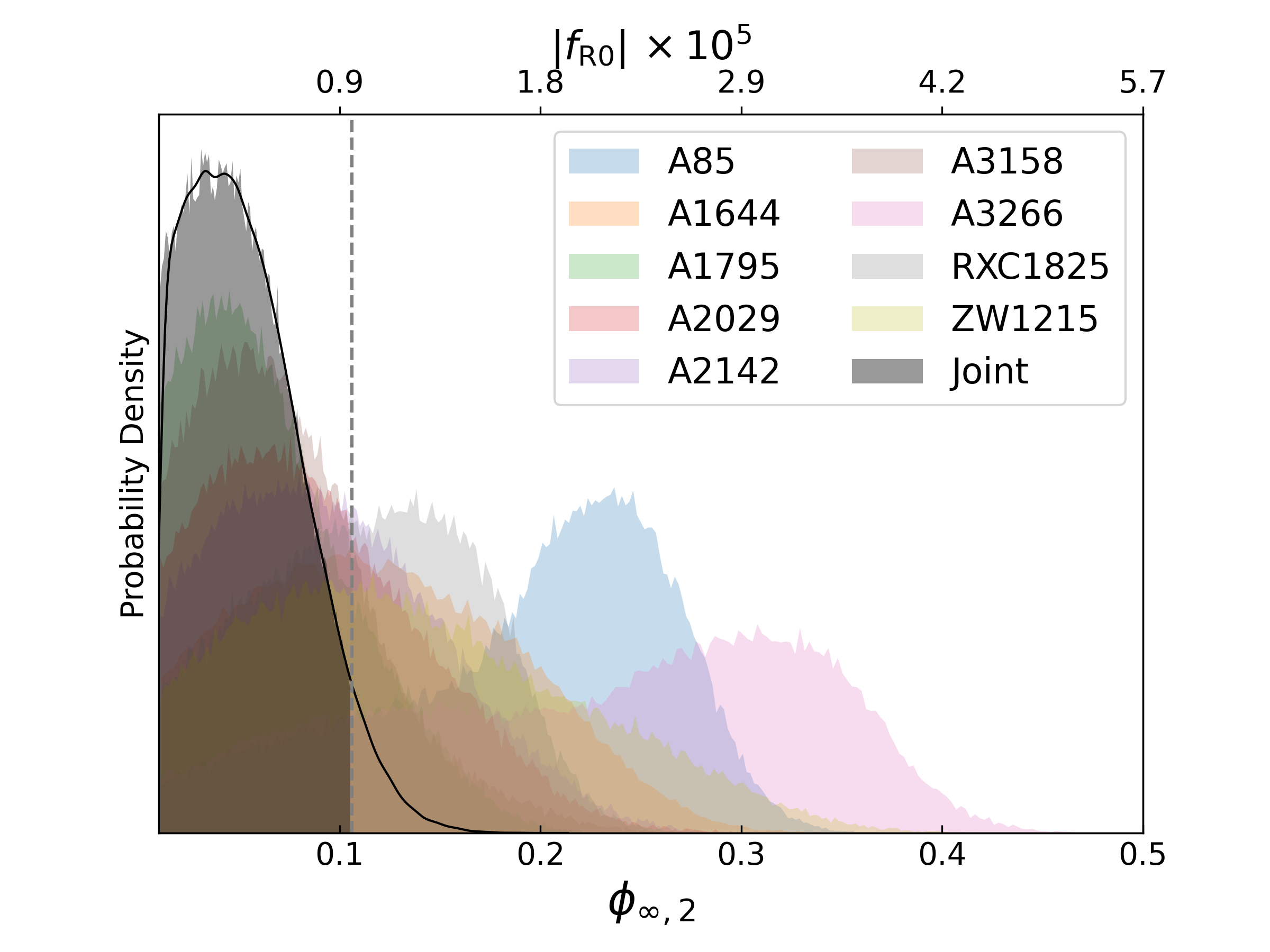}
    \includegraphics[scale=0.46]{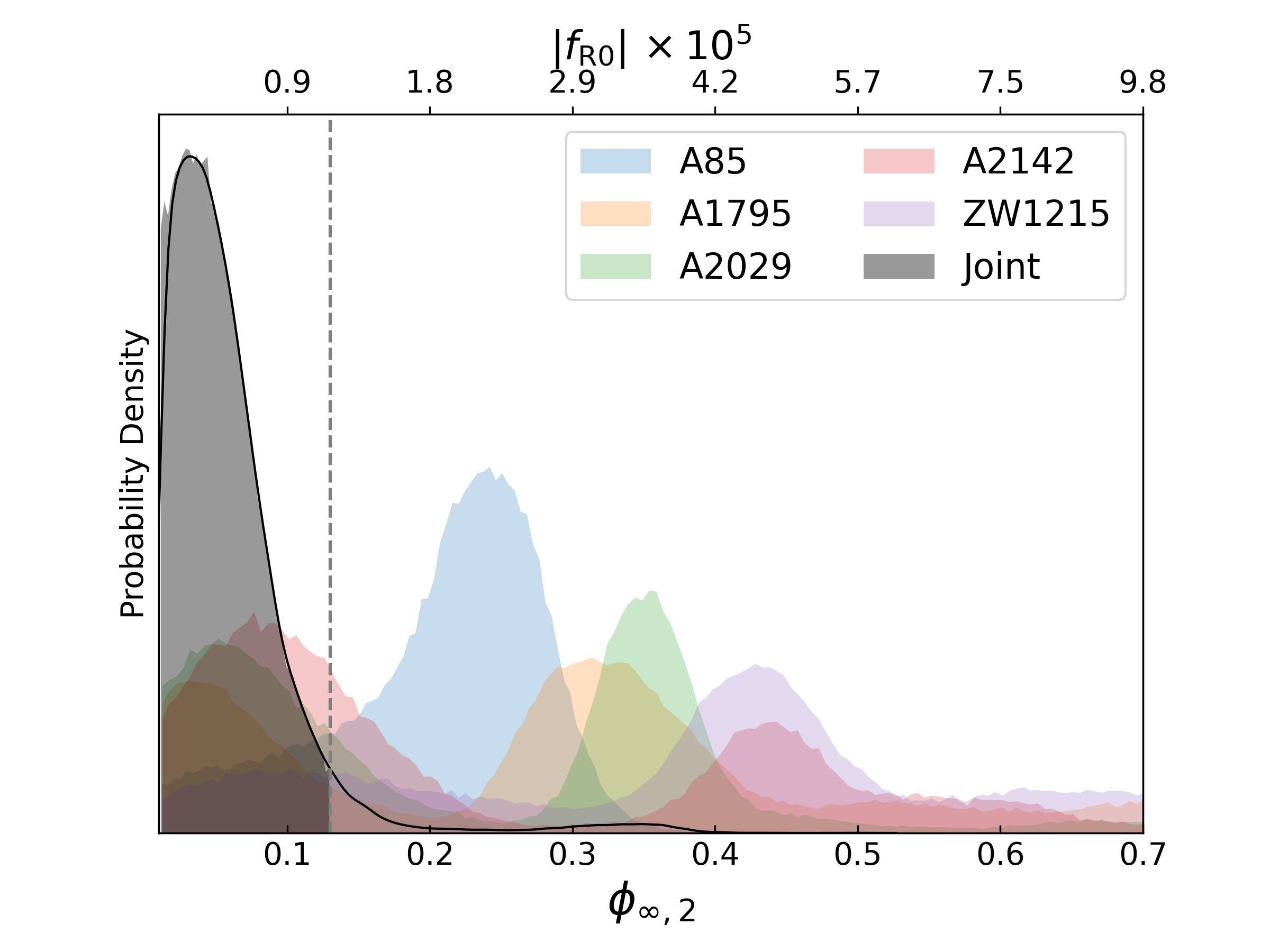}
    
    \caption{Probability distributions for $\phitwo$ ($f_{R0}$ depicted on the top axis) obtained for the specific case of $\beta=1/\sqrt{6}$ for all the clusters within the compilation and the consequent  95\% C.L. regions for the joint analysis (shaded in gray). \textit{Top}: 9 clusters with an internal mass prior. \textit{Bottom}: 5 clusters for which a WL mass prior was available. Please note the difference in the limits of axes in the two figures. }
    \label{fig:test_fr}
\end{figure}

{In the following we use the results of our joint analysis of the chameleon parameter space to place constraints on the background \textit{scalaron} field $f_{R0}$, neglecting the redshift dependence. Starting from the joint posteriors of \Cref{fig:BPR_Joint} and \Cref{fig:MPRNE}, we consider the slice of parameter space $(\btwo,\phitwo)$ for a constant value of $\btwo \sim 0.29$ (i.e. $\beta=1/\sqrt{6}$). We then derive the corresponding posterior $P(\phi_\infty| \beta=1/\sqrt{6})$, which is further related to $f_{R0}$ as a particular case of the chameleon coupling, discussed at the end of Section \ref{Sec:Modell}.
In \Cref{fig:test_fr} we plot the distributions $P(f_{R0})$ for the nine-clusters joint case, assuming internal mass priors (\textit{top}), and for the combination of five clusters with weak lensing priors on $\M$ (\textit{bottom}). The colored areas in gray indicate the regions corresponding to the 95\% C.L. As already mentioned, the mass priors play a fundamental role in breaking the degeneracy among model's parameters. In the case of weak lensing information, the priors are not sufficient to remove all the degeneracy, resulting in a bump in the \textit{scalaron} distribution of individual clusters. Although, the individual clusters in the case of WL priors show a bimodal distribution (except for A85), the joint analysis however is capable to providing a tighter constrain owing to the fact that the second mode in the posterior distribution is spread across the values of the $\phitwo$ and consequently in  $f_{R0}$. As our final constraints, we quote, } 

\begin{equation*}
    |f_{R0}|<9.1\times 10^{-6}\,,
\end{equation*}
at 95\% C.L. for the nine-clusters combined analysis, and similarly
\begin{equation*}
    |f_{R0}|<1.1\times 10^{-5}\,,
\end{equation*}
for the five-clusters weak lensing case.

Within the posteriors of the $\phitwo$ shown in the bottom panel of the \Cref{fig:test_fr}, one could distinguish three distinct contributions (except for A85). The first peak which mainly contributes to joint constraint is marginalized for the $\M$ that does not include the degeneracy, with either of $\{\phitwo, \btwo\}$. While the second peak is an outcome of slightly lower masses, and larger values of the $\C$ parameter, essentially implying a modification to the shape of the mass profile. Finally, the extended tail of the distribution seen for $\phitwo>0.5$ is due to the mild degeneracy between $\{\phitwo, \M\}$, for even lower values of $\M$. However, in the joint analyses the latter two features do not amplify being varied non-overlapping distributions.

Earlier in \cite{Terukina:2013eqa}, a constraint of $|f_{R0}|<6\times 10^{-5}\,$ was set using the hydrostatic and weak lensing observables of the coma cluster at $z = 0.02$, which is even more local in comparison of the redshift range $z \in\{0.04, 0.09\}$ of current X-COP clusters. In this context all the individual clusters in the current analysis provide a tighter constraint (see column 4 of \Cref{tab:Constraints}) and almost an order tighter joint constraint when combining all the data. Our constraint here is also tighter w.r.t to the 58 stacked cluster analysis in \cite{Wilcox:2015kna}, which considers XMM cluster survey and CHFTLenS weak lensing observations in a large redshift range of $z \in \{0.1,1.2\}$. In principle, such a joint analysis considers no cosmological evolution in the field. 
Other works that used galaxy clusters estimated $\fR < 10^{-5}$ (e.g. \cite{Cataneo:2016iav,Pizzuti17}); moreover, \cite{Pizzuti:2020tdl} forecasted a value of $\fR < 10^{-6}$ from the combination of lensing and kinematics mass profile reconstructions of a reasonable sample ($\sim 10$) of clusters. Our analysis confirm that constraints of the same order of magnitude can be reached with combination of high-quality X-ray cluster data with physically-motivated priors in the cluster masses. It is also worth to notice that the bounds derived here are \textit{model-independent}, i.e. no particular functional form for $f(R)$ has been assumed.

\section{Conclusions}
\label{sec:conclusions}

In this paper, we have implemented a formalism, following what done in previous works  \cite{Terukina:2013eqa,Wilcox:2015kna}, to test the chameleon screening in galaxy cluster utilizing the hydrostatic equilibrium data. We have constrained the two parameters describing the Chameleon field, the coupling constant $\beta$ and the value of the field at infinity $\phi_\infty$ by analyzing the dynamics of 9 galaxy clusters in X-COP sample. Chameleon field manifests as a fifth force beyond a certain critical screening radius within a cluster that adds up to the gravitational potential. By performing a full Bayesian analysis of the X-ray-emitting gas pressure and the SZ pressure, along with the electron density, we obtain limits on the aforementioned parameters, essentially excluding a part of the parameter space for this modified gravity scenarios. We summarize the results as follows:

\begin{itemize}
    \item We find that adding a physically-motivated mass prior in our analysis will give a remarkably tight constraints, breaking the degeneracy among model parameters (see also \Cref{App:Mass_prior}). For instance as the main result we present, \Cref{fig:hydrostartic}, where we construct an internal mass prior by eliminating the low mass degenerate regions and use the posterior as a prior in the new MCMC chains, obtaining very tight constraints on $\{\beta , \phi_\infty\}$ compared to previous analysis of Coma Cluster \cite{Terukina:2013eqa} and stacked analysis of XMM Clusters \cite{Wilcox:2015kna}.
    
    \item We have then included additional information on $M_{500}$ from weak lensing analysis in \cite{Herbonnet:2019byy} (see \Cref{fig:MPRNE}). While the results are comparable to what we obtained with the internal mass prior, the weak lensing data are not tight enough to remove the degeneracy completely. 
    
    \item We present our final results in \Cref{tab:Constraints} where we show all the constraints obtained using different mass priors and report a joint constraint eventually on the $\fofr$ class of models presented in \Cref{Sec:Modell}.

    \item We note that the marginalizing or fixing the electron density profile shows no affect on the constraints obtained for the chameleon parameters (see \Cref{fig:A2142_NefixNefree}).
\end{itemize}

It is worth to point out that we have considered only clusters for which the total mass profile (in GR) is well described by the NFW model. Although this choice is physically well-motivated, it is important to explore the effect of different mass parametrization that may better describe the total matter distribution within galaxy clusters in theories of gravity alternative to GR. Indeed the NFW model, despite its wide range of applicability over different scales, might not be the best profile to reproduce the mass distribution of halos in a modified gravity scenario (see e.g. \cite{Corasaniti20} and references therein). In particular, the efficiency of the screening mechanism in chameleon gravity is strictly dependent on the mass model, as one can see from Equation \eqref{eqn:4}. 

Moreover, as shown in \cite{Pizzuti:2020tdl}, the inclusion of kinematics of the member galaxies in clusters to constrain the chameleon parameters can help in reducing the degeneracy even further: both galaxy and ICM move at non-relativistic velocities, following the same gravitational potential. However, the underlying physics is different leading to distinct degeneracy among model parameters. We will investigate these aspects in an upcoming work.

\section*{Acknowledgements}
We thank Paolo Creminelli for useful discussion during the early stages of the project. AL has been supported by the EU H2020-MSCA-ITN-2019 Project 860744 `BiD4BESt: Big Data applications for Black Hole Evolution Studies' and by the PRIN MIUR 2017 prot. 20173ML3WW, `Opening the ALMA window on the cosmic evolution of gas, stars and supermassive black holes'. BSH is supported by the INFN INDARK grant. LP acknowledges support from the Czech Academy of Sciences under the grant number LQ100102101. CB acknowledge support from the COSMOS project of the Italian Space Agency (cosmosnet.it), and the INDARK Initiative of the INFN (web.infn.it/CSN4/IS/Linea5/InDark)

\bibliography{bibliography} 

\newpage
\appendix
\section{}
\label{App:Mass_prior}

\subsection{Effects of Mass prior}

\begin{figure}
    \centering
    \includegraphics[scale=0.46]{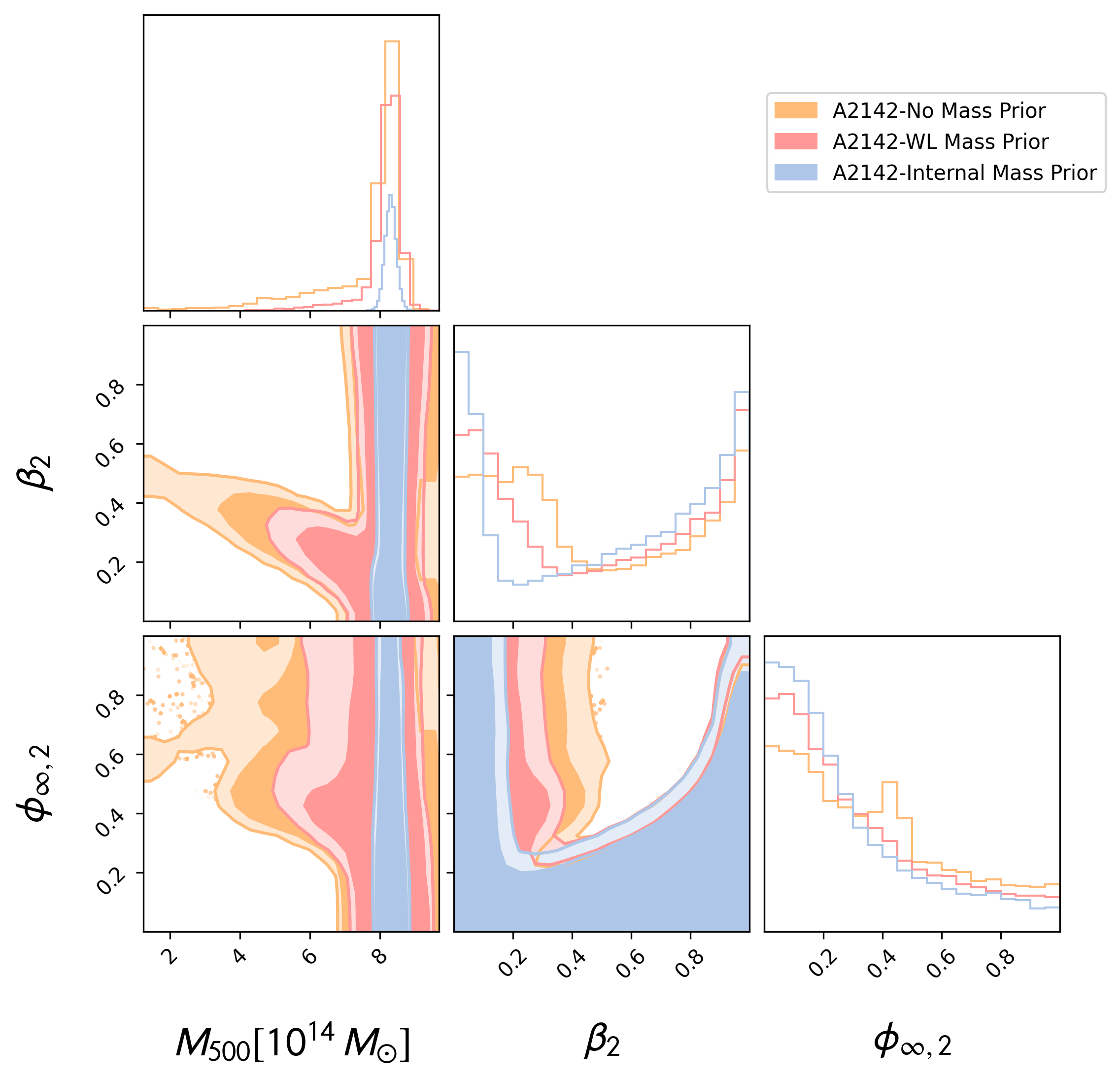}

    \caption{ We show the $95\%$ and $99\%$ C.L. contours for A2142 cluster, where the orange contours represent the ones with no mass prior is taken into account,  the red ones with the weak lensing  mass prior and the blue ones correspond to the internal mass prior case . We notice that the degeneracy region gets reduced when we have an additional mass prior. 
    }\label{fig:compare_3}
\end{figure}

{In this section, we briefly comment on a the different priors choices and systematics due to the electron density data modeling, considering the cluster A2142 as an exemplar. In \cref{fig:compare_3}, we compare the posteriors obtained for these two clusters, with and without the inclusion of the mass priors. The strong degeneracy between the mass of the cluster ($M_{500}$) and the chameleon parameters, can be clearly noticed in the contours shown in blue, deforming the 2-dimensional Gaussian expectation in the $\M$  parameter space. When the WL mass prior is added (shown in red), the degeneracy region shrinks provide more exclusion region in the chameleon parameters. This is completely independent of any analysis choices made and only due do the WL mass prior which is an independent observable, therefore aiding to the constraints. In blue, we show the posteriors when the internal mass prior is considered. As elaborated in \Cref{sec:Results}, this prior is taken from the posterior, when the MCMC analysis is performed with a $\btwo>0.5$ limit. And as expected the mass degeneracy is completely eliminated finding much tighter constrains in the exclusion region. Note that both the mass priors do no modify the constraints of the chameleon parameters for $\btwo >0.5$. }

\begin{figure}
    \centering
    \includegraphics[scale=0.46]{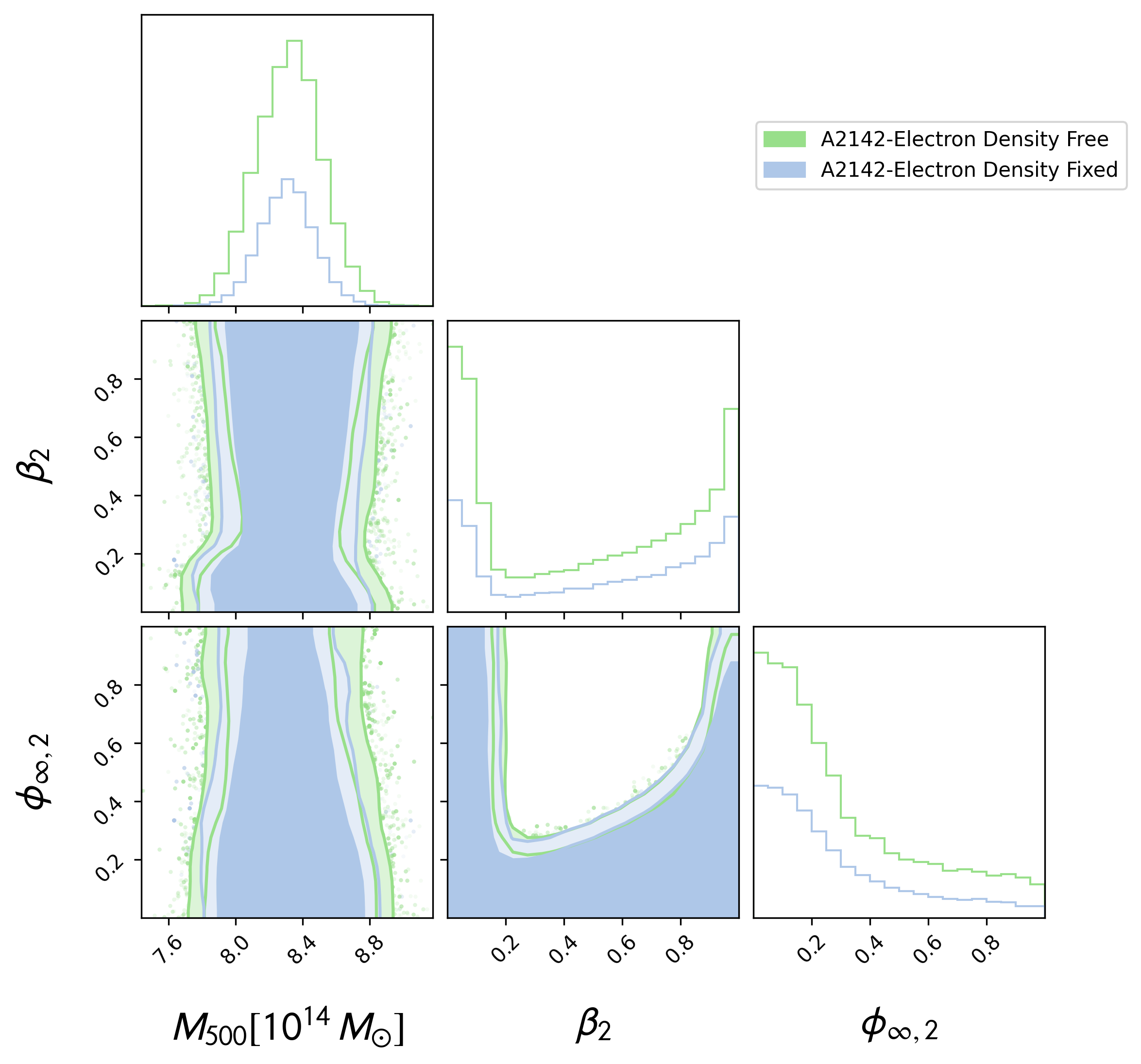}

    \caption{ We show the $95\%$ and $99\%$ C.L. contours for A2142 cluster, wherein the blue contours represent analysis where electron density parameters are fixed. The green contours show the case where the electron density parameters are marginalized upon. 
    }\label{fig:A2142_NefixNefree}
\end{figure}

\subsection{Fixing gas density ($n_e$) profiles }

{In \Cref{fig:A2142_NefixNefree}, we show as an example, the comparison of the contours showing the constrains, when the electron density parameters are allowed to vary in MCMC analysis against the case when they are fixed to the mean values obtained from the former case. We find that the uncertainty in in the electron density parameters does not add to the overall uncertainty in the chameleon parameter space. This can be straightaway understood as there is no expected coupling to the gas density and that the mass profile of the dark matter is modeled via the NFW profile and is assumed to be equivalent to the total mass of the cluster. Noting this as an advantage, we first perform the analysis marginalizing the electron density parameters and latter fixing them to obtain our final results presented in \Cref{sec:Results}. This essentially helps to span the $\{\phitwo, \, \btwo \}$ parameter space effectively in comparison to the the case when all the 10 parameters are allowed to vary, where posteriors might be effected by the sampling methods.} 

\subsection{Alternative weak lensing mass priors}

As noted earlier in \Cref{sec:data,sec:Results}, \cite{Herbonnet:2019byy} provide weak lensing mass estimates using both NFW density profile assumption ($\M^{\rm NFW}$)  and an alternative method, fitting the mean convergence within an aperture radius ($\M^{\rm ap}$), which is independent of the mass profile assumptions. Firstly, we notice that the two masses presented therein are mostly in agreement and utilizing either of them do not change our final constraints, except for the cluster ZW1215 with $\M^{\rm ap} \sim 2\times \M^{\rm NFW}$. We validate that replacing the ZW1215 prior in \Cref{tab:WLpriors} with the higher $\M^{\rm ap}$, considerably improves the exclusion region, however the joint constraint remains unaltered. Therefore, we remain to present our final results with the WL mass priors as the values of $M_{500}$ found assuming the NFW mass profile. 

\end{document}